\documentstyle[12pt,aaspp4]{article}
\lefthead{KALOGERA}
\righthead{ASYMMETRIC SUPERNOVA EXPLOSIONS}
\begin{document}
\begin{center}
To appear in {\em The Astrophysical Journal}
\end{center}
\vspace{1.cm}
\title{Orbital Characteristics of Binary Systems after Asymmetric Supernova 
Explosions} 
\author{Vassiliki Kalogera}
\affil{Astronomy Department, University of Illinois at Urbana-Champaign, \\
1002 West Green St., Urbana, IL 61801. \\
e-mail: vicky@astro.uiuc.edu}
\begin{abstract}
We present an analytical method for studying the changes of the
orbital characteristics of binary systems with circular orbits due to a kick
velocity imparted to the newborn neutron star during a supernova explosion
(SN). Assuming a Maxwellian distribution of kick velocities we derive
analytical expressions for the distribution functions of orbital separations
and eccentricities immediately after the explosion, of orbital separations
after circularization of the post-SN orbits, and of systemic velocities of
binaries that remain bound after the explosion. These distributions of
binary characteristics can be used to perform analytical population synthesis 
calculations of 
various types of binaries, the formation of which involves a supernova
explosion. We study in detail the dependence of the derived distributions 
on the kick 
velocity and the pre-SN characteristics,   
 we identify all the limits imposed on the post-SN
orbital characteristics, and we discuss their implications for the population of
X-ray binaries and double neutron star systems. 
We show that large kick velocities do not necessarily
result in large systemic velocities; for typical
X-ray binary progenitors the maximum post-SN systemic velocity is comparable to the relative orbital velocity prior
to the explosion. We also find that, unless accretion-induced collapse 
is a viable formation channel, X-ray binaries in globular clusters
have most probably been formed by stellar dynamical interactions only, and not 
directly from primordial binaries.
\end{abstract}
\keywords{stars: binaries -- stars: neutron -- stars: supernovae -- X-rays: binaries}

\section{INTRODUCTION}

Studies of the radio pulsar population (e.g., Gunn \& Ostriker\markcite{G70} 1970; Helfand 
\& Tademaru\markcite{H77} 1977; Harrison, Lyne \& Anderson\markcite{H93} 1993; Lyne 
\& Lorimer\markcite{L94} 1994) have shown that pulsars move in the Galaxy 
with very high space velocities,  
ranging from $20$ to $2000$\,km\,s$^{-1}$, and that their galactic
distribution  has a large scale height, of the order of $1$\,kpc. The origin
of these  high velocities is often attributed to a kick velocity imparted to
the  neutron star at the time of the supernova explosion. Early studies by 
Dewey \& Cordes\markcite{D87} (1987) and Bailes\markcite{B89} (1989) 
concluded that the mean magnitude of 
the kick velocity is of the order of  
$100-200$\,km\,s$^{-1}$. In a more recent study,  
which takes into account new measurements of pulsar proper motions, a new 
electron density model, and a selection effect against fast pulsars,  
Lyne \& Lorimer (1994) found  
the mean kick velocity to be $\sim 450\pm 90$\,km\,s$^{-1}$. 
Additional observational evidence in support of a kick 
velocity imparted to neutron stars at birth are related to the existence of a
high--velocity population of O,\,B runaway stars (e.g., Stone\markcite{S91} 1991), as well as 
to supernova remnant -- pulsar associations, studies of which yield 
kick velocities up to 
$2000$\,km\,s$^{-1}$ (Caraveo\markcite{C93} 1993; Frail, Goss, \& Whiteoak\markcite{F94} 1994). 

In contrast to Dewey \& Cordes (1987), Iben \& Tutukov\markcite{I96} (1996) have recently 
concluded that the hypothesis of natal kicks imparted to neutron stars is 
unnecessary. They have found 
that the transverse
velocity distribution of pulsars in the solar neighborhood, as well as that
of O,\,B runaway stars, massive X-ray binaries, and double neutron stars, 
can be explained by the recoil velocity due to symmetric supernova explosions. 
However,
they reach this conclusion by assuming (i) that all stars are 
members of binary 
systems and (ii) that neutron stars  
formed by massive single stars (formed only by mergers) or 
in wide binary systems rotate too slowly to become radio pulsars.  
Although, their results are
marginally consistent (mean predicted velocities are
$\sim 100-150$\,km\,s$^{-1}$) with the old pulsar distance scale
(Harrison et al. 1993), they are not consistent with the more recent results 
of Lyne \& Lorimer
(1994). 

Over the years, several theories have been put forward in an effort to explain 
the origin of kick velocities 
(e.g., Harrison \& Tademaru\markcite{H75} 1975; Chugai\markcite{C84} 1984; 
Duncan \& Thomson\markcite{D92} 1992; Herant, Benz \& Colgate\markcite{H92} 1992; 
Janka \& M\"{u}ller\markcite{J94} 1994; Burrows, Hayes, \& Fryxell\markcite{F95} 1995; 
Burrows \& Hayes\markcite{H96} 1996). 
Even a small 
asymmetry during the collapse of 
the core can give a kick to the remnant of the explosion. The asymmetry 
may be related either to neutrino emission or to mass ejection during the 
supernova, and may be 
caused by the magnetic field or rotation of the collapsing core, or 
by hydrodynamic instabilities, such as Rayleigh-Taylor or convective motions. 
In any case, the mechanism responsible for the kick 
velocity is still not well understood, and it appears that fully 
three-dimensional numerical simulations of the core collapse will be 
required in 
order to settle this issue. 

Several authors have previously studied the effect of an 
asymmetric supernova explosion on binary parameters, 
focusing on various aspects of the 
problem. 
Early work by Flannery \& van den Heuvel\markcite{F75} (1975), 
Mitalas\markcite{M76} (1976), Sutantyo\markcite{78} (1978), and 
Hills\markcite{H83} (1983) addressed the problem of deriving expressions 
of post-SN orbital characteristics for a specific kick velocity for both 
circular and eccentric pre-SN orbits. They also derived survival probabilities
for kick velocities of constant magnitude and random direction. The one-to-one 
link between pre-SN and post-SN parameters is broken when kick velocities 
are allowed to have a distribution over both magnitude and direction, in 
which case there exists a distribution of post-SN characteristics, even for 
pre-SN binaries with specific orbital parameters. 
Wijers, van Paradijs, \& van den Heuvel\markcite{W92} (1992) were the first to 
address this problem. They derived an analytic expression for the 
distribution of post-SN orbital separations and eccentricities only, which 
however was also convolved with   
a distribution of pre-SN orbital separations. More recently  
Brandt \& Podsiadlowski\markcite{B95} (1995) addressed the same problem using 
numerical methods (Monte Carlo simulations). The resulting 
distributions are again convolved with pre-SN period distributions, and, in 
this case, are 
calculated only for specific stellar masses, in an effort to compare them with 
observation. 
Because these
distributions are calculated numerically, information about the
allowed ranges of post-SN characteristics, the shape of multi-dimensional
distributions, and their dependence on the pre-SN and kick-velocity
characteristics is limited.
Our purpose here is to derive analytical expressions of various post-SN 
characteristics for the realistic case of kick velocities with a distribution 
in both magnitude and direction. The derived distributions are general, 
and apply to any circular binary systems that experience asymmetric 
supernova explosions. 

The study presented in this paper has been motivated by our interest in 
performing population synthesis calculations for low-mass X-ray binaries. 
Monte Carlo techniques have been widely used in such calculations 
modeling various kinds of binary systems (e.g., Dewey \& Cordes 1987; 
de Kool\markcite{92} 1992; Romani\markcite{R92}
 1992).  
Another method is based 
in creating a multi-dimensional grid of initial binary parameters and tracing 
the evolution of systems through a sequence of evolutionary stages for each 
set of initial parameters (Kolb\markcite{93} 1993; Iben, Tutukov, \& Yungel'son\markcite{I95} 1995 and 
references therein). Both of these numerical methods have the same 
problem: although the goal is to calculate the characteristics of the final 
population, the sampling procedure is applied on the initial population, 
and therefore it is possible that the final population is under-sampled, 
even if the sampling of the primordial population appears adequate. 
Another problem with both methods is related to statistical accuracy: the 
initial sets of parameters cover a wide range, of which only a small part is 
populated by progenitors of interest, especially in the case of X-ray binaries, 
which have very small birth rates; therefore, with these methods 
it is necessary to study a very 
large number of primordial binaries in order for a statistically significant 
number of systems to survive. Both of these problems are absent in population 
synthesis calculations performed analytically, where distributions of 
primordial binaries over orbital characteristics are transformed through a 
sequence of evolutionary stages, using Jacobian transformations. In this way, 
the regions in parameter space populated by the progenitors of interest are 
identified and the final population is calculated directly. This method has 
been formulated and applied to the study of cataclysmic binaries by 
Politano\markcite{P96} (1996) (see also Politano, Ritter, \& Webbink\markcite{P89} 1989; Politano \& 
Webbink\markcite{P90} 
1990). Apart from the absence of the problems discussed above, the analytical 
method has the additional advantage that the shape of final distributions 
is calculated exactly, revealing fine details and subtle features, such as 
sharp peaks, infinities, or definite limits imposed on the final parameters. 
Also, 
the various dependences of these parameters and their distributions on the 
initial parameters can be identified and studied in detail. 

In order to use the analytical method in population synthesis of neutron-star 
binaries, it is necessary to develop an analytical tool for the modeling of 
asymmetric supernova explosions. Our purpose in this paper is to 
present such a method 
based on Jacobian transformations for computing analytically the probability 
distributions of several orbital 
characteristics of post-SN binaries. These distributions include (\S\,2) 
the orbital separations and eccentricities immediately 
after the supernova explosion, the circularized orbital separations, and 
(\S\,3) the  
systemic velocities. They are derived for circular pre-SN 
orbits and for kick velocities that 
are randomly distributed not only in direction but 
also in 
magnitude (Maxwellian distribution). The derived expressions can be used in 
synthesis calculations of 
any kind of binaries  that experience supernova explosions during their 
evolution. 
The analytical character of the derivation enables us to perform detailed 
parameter studies and identify those characteristics of the kick velocities or 
the pre-SN binaries that govern the behavior of the post-SN distribution 
functions. 
Expressions for the limiting cases of very large or very small kick 
velocities relative to the pre-SN orbital velocities are 
also derived. We identify the limits imposed 
on the post-SN parameters and discuss their physical interpretation. In 
addition, 
we calculate 
survival probabilities (\S\,4) 
as functions of the pre-supernova (pre-SN) orbital 
characteristics and the mean kick velocity. Finally, we examine 
several implications of our results (\S\,5)  
for the 
progenitors of high- and low-mass X-ray binaries, double 
neutron stars, and their populations in globular clusters.
A list of the symbols used throughout the paper is given in Appendix A. 
The study of a few special cases is included in Appendices B and C.   
     
\section{POST-SUPERNOVA ORBITS}

We assume that the binary orbits prior to the supernova explosion are 
circular, and that the kick velocities follow  
a Maxwellian distribution. 
The first of these two assumptions is unlikely to be violated in the case 
of systems that have emerged from a common-envelope phase, or have experienced  
(semi-)conservative mass transfer. Orbital eccentricities may be important for 
binaries with components that have not interacted prior to the supernova 
explosion, although it is still possible that circularization has occurred 
during 
their main-sequence evolution (see Portegies-Zwart \& Verbunt\markcite{P96} 1996). 
The second assumption, that of a Maxwellian distribution of kick velocities, 
we adopt in the absence
of an adequate theoretical understanding of their origin. It is 
conceivable that the direction of the kick is affected by the 
kinematical or rotational properties of the collapsing core, but any 
correlation between the kick direction and the orbital rotational axis or 
the orbital velocity has yet to be established. In principle, a method 
like that 
described here can be used with any distribution of kick velocities, 
although the 
derivational details will be different.  
 
In most of our calculations the interaction between the expanding 
supernova shell and the companion to the exploding star has been ignored. 
According to Fryxell \& Arnett (1981), this interaction is generally weak, 
especially 
in the case of low-mass X-ray binaries (LMXBs) and double neutron stars, 
where the solid angle intercepted 
by the companion is very small unless the orbital separation is also small 
(see Appendix A and Romani 1992). 
This effect may be more important for high-mass X-ray binaries (HMXBs), 
although their orbits are much wider. Nevertheless, for completeness 
we have repeated 
the calculation of the probability distribution of circularized orbital 
separations, including also the effect of the impulse velocity (see Appendix A). 

\subsection{Non-Circularized Orbits}

In this section we derive the distribution of post-SN binary systems over 
orbital separations and eccentricities immediately 
after the supernova explosion,
 by transforming 
the distribution of kick 
velocities imparted to the neutron star into a distribution over 
other binary parameters of interest. 

 We adopt a reference frame centered on the 
exploding star mass, $M_1$, just prior to the supernova. The 
companion mass, $M_2$, 
is chosen to be at rest and the exploding star to move in a circular 
orbit with separation $A_i$. The x-axis lies along the line 
connecting the centers of mass 
of the two stars, pointing from $M_2$ to $M_1$. The y-axis lies parallel
to the direction of  
the pre-SN orbital velocity $V_r$ of $M_1$ relative to its companion.
The  z-axis completes a right-handed
orthogonal system (Figure 1).

Two of the parameters characterizing the post-SN binaries are the orbital 
separation, $A_f$, and the eccentricity, $e$. We use the energy and angular 
momentum equations for eccentric orbits to relate these two parameters to the 
three components of the kick velocity. 
The supernova explosion, mass loss, and kick are assumed 
to be instantaneous. 
In general, for
two stars 
with masses $M_a$ and $M_b$ in an orbit with orbital separation $A$ and
eccentricity
$e$, their relative velocity $V$ at a distance $r$ is given by:
\begin{equation}
V^2~=~G~(M_a+M_b)~\left(\frac{2}{r}-\frac{1}{A}\right)\mbox{,}
\end{equation}
and the  specific angular momentum of the system is:
\begin{equation}
\vert \vec{r}\times \vec{V}\vert ^2~=~G~(M_a+M_b)~A~(1-e^2).
\end{equation}
We can also apply the above two equations in the case of the post-SN binary, 
for which
 $\vec{V}=(V_{kx},
V_{ky}+V_r,V_{kz})$ and $\vec{r}=(A_i,0,0)$. The 
orbital separation, $A_f$, and the eccentricity, $e$, of the post-SN orbit 
are thus related 
to the components of the kick velocity, $V_{kx}$, $V_{ky}$, and $V_{kz}$ 
by the expressions: 
\begin{eqnarray}
A_f & = & G(M_{NS}+M_2)\left[\frac{2G(M_{NS}+M_2)}{A_i}-V_k^2-V_r^2-2V_{ky}V_r\right]^{-1} \\
1~-~e^2 & = & \frac{(V_{kz}^2+V_{ky}^2+V_r^2+2V_{ky}V_r)A_i^2}{G(M_{NS}+M_2)A_f}\mbox{,}
\end{eqnarray}
where $M_{NS}$ is the gravitational mass of the neutron star and 
 $V_k$ is the magnitude of the kick velocity. 

The third independent parameter describing the post-SN state of the binary is
the  orientation of the eccentric orbit relative to the pre-SN 
orbital plane. The plane of the binary  
orbit is altered due to the z-component of the kick velocity.  Since the explosion is assumed to be 
instantaneous, the position of the two stars just before and just after the 
supernova remains unchanged. In our reference frame (Figure 1) the two stars 
lie along the x-axis, and therefore 
the intersection of the two orbital planes must coincide with the x-axis. 
The angle, $\theta$, between the pre- and post-SN
orbital planes, is equal to the one between the  relative velocity just before
the explosion, $\vec{V_r}=(0,V_r,0)$, and the  projection  
of the relative velocity just after the explosion onto the $y-z$ plane, 
$\vec{V}_{yz}=(0,V_{ky}+V_y,V_{kz})$. Hence:
\begin{equation}
cos\theta ~=~ \frac{\vec{V}_{yz}\cdot \vec{V_r}}{\vert \vec{V_r} \vert ~ \vert 
\vec{V}_{yz}\vert }~=~ 
\frac{V_{ky}+V_r}{\left[(V_{ky}+V_r)^2+V_{kz}^2\right]^{1/2}}.
\end{equation}

For convenience, we rewrite equations (3),\,(4),\,(5) 
in dimensionless form using
the  following 
definitions:
\begin{eqnarray}
\alpha & \equiv & \frac{A_f}{A_i}\mbox{,} \\
\beta & \equiv & \frac{M_{NS}+M_2}{M_1+M_2}\mbox{,} \\
v_{kj} & \equiv & \frac{V_{kj}}{V_r}\mbox{,}
\end{eqnarray} 
where $j$ can be $x$, $y$, or $z$. 
The dimensionless post-SN orbital separation 
$\alpha$, the eccentricity $e$, and the angle $\theta$ between pre- and 
post-SN orbital planes can be 
expressed as functions 
of $v_{kx}$, $v_{ky}$, $v_{kz}$, and $\beta$:
\begin{eqnarray}
\alpha & = & \frac{\beta}{2\beta-v_{kx}^2-(v_{ky}+1)^2-v_{kz}^2} \\
1~-~e^2 & = & \frac{v_{kz}^2+(v_{ky}+1)^2}{\beta^2}~\left[2\beta-v_{kx}^2-(v_{ky}+1)^2-v_{kz}^2\right] \\
\cos \theta & = & \frac{v_{ky}+1}{\left[v_{kz}^2+(v_{ky}+1)^2\right]^{1/2}}. 
\end{eqnarray}

To obtain the distribution of binaries over post-SN characteristics we use 
the Jacobian transformation of the kick velocity distribution. 
The distribution functions for each of the three components of the kick 
velocity 
are assumed to be Gaussian:   
\begin{equation}
p(v_{kx},v_{ky},v_{kz})~=~\prod_{j} \frac{1}{\sqrt{2\pi \xi ^2}}~  
\exp \left(-\frac{v_{kj}^2}{2\xi ^2}\right)\mbox{,}
\end{equation}
where $\xi \equiv \sigma /V_r$, and $\sigma$ is the velocity dispersion of 
each of the one-dimensional Gaussian distributions. 
Using equations (9), (10), and (11) we obtain:
\begin{eqnarray}
g(\alpha,e,\cos \theta) & = & 
\left(\frac{\beta}{2\pi\xi^2}\right)^{3/2}
\frac{2~e}{\left[\alpha (1-e^2)\right]^{1/2}}~\left[\left(\alpha-\frac{1}{1+e}
\right)\left(\frac{1}{1-e}-\alpha\right)\right]^{-1/2}\nonumber \\
 &  & \times \exp \left[-\frac{1}{2\xi^2}
\left(\beta~\frac{2\alpha-1}{\alpha}+1\right)\right]\nonumber \\
 &  & \times \exp \left[\frac{\left(\beta~\alpha ~(1-e^2)\right)^{1/2}}{\xi^2}\cos \theta\right]~
\left(1-\cos ^2\theta\right)^{-1/2}. 
\end{eqnarray}

In general, we are not interested in  
the orientation of the 
post-SN orbital plane. Therefore, we integrate $g(\alpha,e,\cos \theta)$  
over all orientations, and obtain for the distribution over orbital 
separations and eccentricities:   
\begin{eqnarray}
G(\alpha,e) & = & \left(\frac{\beta}{2\pi\xi^2}\right)^{3/2}
\frac{2\pi e}{\left[\alpha (1-e^2)\right]^{1/2}}~
\left[\left(\alpha-\frac{1}{1+e}
\right)\left(\frac{1}{1-e}-\alpha\right)\right]^{-1/2}\nonumber \\
 &  & \times \exp \left[-\frac{1}{2\xi^2}
\left(\beta~\frac{2\alpha-1}{\alpha}+1\right)\right]~I_o(z),
\end{eqnarray}
where
\begin{displaymath}
z~\equiv~
\frac{\left(\beta~\alpha~(1-e^2)\right)^{1/2}}{\xi^2},
\end{displaymath}
and $I_o$ is the modified Bessel function of zeroth order.  
The above expression has two singularities, at 
$\alpha=1/(1\pm e)$, which correspond to the special case of the velocity 
of the newborn neutron star
being restricted in the y-z plane,  
$V_{kx}=0$. We can see this by using equations (9),\,(10) to obtain:
\begin{equation}
v_{kx}^2~=~\frac{\beta}{\alpha}~(1-e^2)~\left(\alpha-\frac{1}{1+e}\right)~
\left(\frac{1}{1-e}-\alpha\right)
\end{equation}
In the singular cases, the distribution of kick velocities becomes 
two-dimensional,
and there are only two independent variables describing the post-SN state:
$\cos \theta$ and $e$ (or $\alpha$). The corresponding distributions are derived
in Appendix B. 
 
Scrutiny of equation (14) indicates that post-SN binaries populate only a 
restricted
area of the 
$\alpha - e$ plane, independent of the 
orbital characteristics of the pre-SN systems. Acceptable values for $\alpha$ 
span a range from $1/(1+e)$ to $1/(1-e)$, limits which were first identified by 
Flannery \& van den Heuvel (1975).   
Since the  post-SN orbit must include the position of the two stars just
prior to  the explosion,  
the post-SN
orbital separation, $A_f$, cannot be smaller  than half of the pre-SN
separation, $A_i$. 

In addition to remaining bound,  
the two stars in the post-SN binary must avoid physical collision, which 
would probably lead to a merger. 
The closest distance between
the two stars must at a minimum exceed the sum of their radii:
\begin{equation}
A_f~(1-e)~>~R_{NS}+R_2\simeq R_2\mbox{,}
\end{equation}
or
\begin{equation}
\alpha~(1-e)~>~\frac{R_2}{A_i}\equiv c.
\end{equation}
This condition sets a lower limit on $\alpha$, 
$\alpha > c/(1-e)$, 
or an upper limit on $e$, $e<1-c/\alpha$. 
The complete set of limiting curves on the $\alpha-e$ parameter space, 
for a range of different values of $c$,  
is shown in Figure 2. 
Nevertheless, 
we will set $c=0$ for simplicity in the following discussion, 
returning at the very end of this section 
to comment on the effect of a non-zero value of $c$. 

A two-dimensional distribution over $\alpha$ and $e$ (eq.\,[14]) 
is shown in
Figure 3  for the specific choice of 
$\beta =0.6$ and $\xi =1$.  The behavior of the distribution is
dominated by the square root term that  appears in equation (14). This 
term becomes
equal to zero along the $\alpha (1\pm e)=1$ curves.   
Variation  of the values of $\beta$ and
$\xi$ affects only the normalization of the  distribution, and not its
qualitative shape.   

The distribution of post-SN systems over eccentricity, ${\cal J}(e)$,  can be
found by integrating 
$G(\alpha,e)$ over $\alpha$.
Sample distributions of post-SN systems over $e$ are plotted in 
Figures 4a and 4b 
for different values of $\beta$ and $\xi$. 
For $\beta \geq 0.5$ (Figure 4a), less than half 
the total mass of the pre-SN system is lost, and the binary would remain bound 
in the case of a symmetric explosion (no kick imparted to the neutron star). 
In the limit that $\xi \rightarrow 0$, the distribution over $e$ sharply peaks 
at $e\rightarrow (1-\beta)/\beta$, which is the eccentricity of the post-SN 
orbit if the explosion were symmetric (see, for example Verbunt\markcite{V93} 1993). As 
$\xi$ increases, and the kick velocity becomes comparable to the relative 
orbital velocity of the stars prior to the supernova explosion, the 
distribution becomes broader, and then declines uniformly for $\xi \gtrsim 1$. 
When $\beta < 0.5$ (Figure 4b), binaries would be disrupted in the 
absence of any kick velocity. In the limit that $\xi \rightarrow 0$, 
few systems 
survive generally with very high eccentricities ($e\rightarrow
1$). As
$\xi
$ increases, ${\cal J}(e)$ grows until $\xi \sim 1-\sqrt{\beta}$, 
then declines.  
For $\xi \gg 1$, ${\cal J}(e)$ converges to the same asymptotic form 
regardless of whether $beta > 0.5$ or not:
\begin{equation}
\lim_{\xi \rightarrow \infty}{\cal J}(e) = 4\pi~\left(\frac{\beta}{2\pi\xi^2}\right)^{3/2}~\frac{e}{\sqrt{1+e}}~
K(p),
\end{equation}
where
\begin{displaymath}
p\equiv \sqrt{\frac{2e}{1+e}},
\end{displaymath}
and $K(p)$ is the complete elliptic integral with the following series 
representation:
\begin{displaymath}
K(p) = \frac{\pi}{2}\left\{1+\left(\frac{1}{2}\right)^2p^2+\left(\frac{1\cdot 3}
{2\cdot 4}\right)^2p^4+\mbox{...}+\left[\frac{(2n-1)!!}{2^nn!}\right]^2p^{2n}+
\mbox{...}
\right\},
\end{displaymath}
where $n$ is a positive integer. 

To  derive the distribution over orbital separations, ${\cal G}(\alpha)$, we 
integrate over eccentricities, $e$. 
Plots of the distribution over $\alpha$ are given in 
Figures 5a and 5b for a set of different $\beta$ and $\xi$.
When $\beta \geq 0.5$ and $\xi\rightarrow 0$ (Figure 5a), conditions are 
similar to those in a symmetric supernova. The distribution is 
narrow and peaks at $\alpha\rightarrow\beta/(2\beta -1)$, 
which is the post-SN orbital separation in the absence of any kicks 
(see Verbunt 1993). 
As $\xi$ increases the distribution broadens and peaks at orbital separations 
smaller than that before the explosion.  
For 
$\beta < 0.5$ (Figure 5b) 
and small kick velocities, the probability of disruption is 
very high and the binaries that survive have very large orbital separations. 
For higher values of $\xi$, more systems are able to reduce their 
energy and remain bound with smaller orbital separations.  
When $\xi \gg 1$, ${\cal G}(\alpha)$ converges to the asymptotic form:
\begin{equation}
\lim_{\xi \rightarrow \infty}{\cal G}(\alpha) = 2\pi~\left(\frac{\beta}{2\pi\xi^2}\right)^{3/2}~
\frac{\sqrt{2\alpha -1}}{\alpha^{5/2}}, 
\end{equation}
where $\alpha \geq 1/2$. We note that for $\xi \gg 1$, both ${\cal G}(\alpha)$
and ${\cal J}(e)$ assume forms which, apart from their normalization do not 
depend on either $\xi$ nor on $\beta$. 

In the preceding discussion, we have not accounted for the possibility that
the two stars may collide after the supernova explosion. As a 
consequence of this last constraint the parameter space in
$\alpha$ and $e$ is further restricted, and the integrated distributions  of
orbital separations and eccentricities are altered (see Figures 6a and 6b).
It is evident that 
 the
survival probabilities decrease dramatically if $c$ is not very small 
($c\gtrsim 0.01$), 
especially for large  eccentricities and
for orbital separations larger than that of the pre-SN binary.  

\subsection{Circularized Orbits}

Tidal interaction between the binary members leads to circularization of the
post-SN  orbit, on a time scale that depends 
on the characteristics both of the eccentric orbit and of the companion to 
the neutron star. Setting aside the question of the relevant time scales
we can calculate the distribution of post-SN systems over orbital separation, 
$A_c$,  after circularization has been achieved. 

During the circularization process, orbital energy, $E$, is dissipated while 
orbital angular momentum, $J$, is conserved.
We define the dimensionless quantities:
\begin{eqnarray}
j^2 & \equiv & \left(\frac{J}{J_o}\right)^2 ~ = ~ \alpha~(1-e^2), \\
\epsilon & \equiv & \frac{E}{E_o} ~ = ~ \frac{1}{\alpha},
\end{eqnarray} 
where $J_o^2\equiv G\,A_i\,M_{NS}^2M_2^2/(M_{NS}+M_2)$ and $E_o\equiv 
-G\,M_{NS}M_2/(2A_i)$. 
From conservation of orbital angular momentum we find for the orbital 
separation, $A_c$, 
of the circularized orbit:
\begin{equation}
j^2~=~\frac{A_c}{A_i}\equiv \alpha_c.
\end{equation}
Using equations (20), (21), and (22) we transform $G(\alpha,e)$ into the  
distribution of post-SN circularized systems over $\alpha_c$ and $\epsilon$:
\begin{equation}
H(\alpha_c,\epsilon)  =  \pi \left(\frac{\beta}{2\pi\xi^2}\right)^{3/2}~
\exp \left(-\frac{2\beta +1}{2\xi^2}\right)~I_o\left[\frac{\sqrt{\beta 
\alpha_c}}{\xi ^2}\right] 
\frac{\exp \left(\frac{\beta}{2\xi ^2}\epsilon\right)}
{\sqrt{2-\alpha_c-\epsilon}}\mbox{,}  
\end{equation}
\begin{displaymath}
\hspace{4cm} \mbox{where}~~~(2-\alpha_c-\epsilon)>0. 
\end{displaymath} 

The circularized post-SN systems are characterized 
by only one parameter, the orbital separation. In order to find their
distribution over $\alpha_c$ we need to integrate $H(\alpha_c,\epsilon)$ 
over the dimensionless orbital energy $\epsilon$. The limits of integration 
are found by considering all the constraints that viable post-SN systems 
must satisfy. 

An upper limit to $\epsilon$ is set by the geometrical constraint,  
that the post-SN eccentric orbit must include the position 
of the stars prior to the supernova explosion (see eq.\,[23]),   
\begin{equation}
\epsilon ~<~2 - \alpha_c. 
\end{equation}
The second constraint is that the post-SN system must be bound, and  
hence its orbital energy $E$ must be negative. Since $E_o$ has been 
defined to be negative, we obtain a lower limit for $\epsilon$:
\begin{equation}
\epsilon > 0. 
\end{equation}
An additional lower limit is set by the need to avoid a physical collision.  
This condition is expressed 
as a lower limit on $\alpha$, i.e., $\alpha > c/(1-e)$, such that the
periastron  distance in the 
eccentric orbit exceeds the 
radius of the companion to the neutron star. Using equations 
(20), (21), and (25),  
we can rewrite this condition as:
\begin{equation}
\epsilon~>~\frac{2c-\alpha_c}{c^2}. 
\end{equation}

By checking for consistency ($\epsilon_{max}>\epsilon_{min}$) among the above 
limits, 
we find that from all possible values of $\alpha_c$ only a small range is 
acceptable 
for post-SN circularized orbits:
\begin{equation}
\frac{2c}{1+c}~<~\alpha_c~<~2.
\end{equation}
 
The post-SN systems may be divided into groups depending on the value of 
$\alpha_c$. 
Systems with $\alpha_c\geq 2$ become unbound ($\epsilon$ becomes negative). 
Systems with $\alpha_c\leq 2c/(1+c)$ are bound, but all lead to a merger of the 
two stars. For $2c<\alpha_c<2$, systems are bound and they all avoid collision. 
In this case, $\epsilon_{min}=0$ and $\epsilon_{max}=2-\alpha_c$. Finally, 
systems with $2c/(1+c)<\alpha_c<2c$ are also bound, but a fraction of them 
merge. For this reason the range of acceptable energies is further constrained:
$\epsilon_{min}=(2c-\alpha_c)/c^2>0$. Clearly, unlike the case 
of a
 symmetric explosion, where the 
supernova always results in an expansion of the orbit (Verbunt 1993), 
the separation 
of the circularized orbit after an asymmetric explosion 
may become smaller than the pre-SN separation.
However, although the lower limit on $\alpha$ is 
extended to values smaller than 
unity due to the kick velocity, the upper limit to 
the post-SN circularized orbital separation remains twice the 
pre-SN separation. 

To obtain the distribution of post-SN binaries over $\alpha_c$ we 
integrate over $\epsilon$. 
This integration can be performed analytically, yielding:  
\begin{equation}
{\cal H}(\alpha_c)~=~ \left(\frac{\beta}{2\xi^2}\right)~
\exp\left(\frac{-\beta \alpha_c+1}{2\xi^2}\right)~I_o\left(\frac{\sqrt{\beta\alpha_c}}{\xi^2}
\right)~\mbox{erf}\left(z_o\sqrt{\frac{\beta}{2\xi^2}}\right),
\end{equation} 
where:
\begin{eqnarray}
\mbox{erf}(x_o) & \equiv & \frac{2}{\sqrt{\pi}} \int_0^{x_o}\mbox{e}^{-x^2}~
dx\mbox{,} \nonumber \\ 
z_o & = & \sqrt{2-\alpha_c-\frac{2c-\alpha_c}{c^2}}\mbox{,}~~~~~~~~~
\frac{2c}{1+c}<\alpha_c<2c \nonumber \\
 & = & \sqrt{2-\alpha_c}\mbox{,}\hspace{3.7cm} 2c\leq\alpha_c<2. \nonumber
\end{eqnarray}

The behavior of ${\cal H}(\alpha_c)$ is dictated by the values of the two parameters 
$\beta$ and $\xi$. Using the asymptotic forms of the error function and the 
modified Bessel function in the two limits that the r.m.s. kick velocity is 
much larger or much smaller than the relative orbital velocity in the 
pre-SN orbit we obtain:
\begin{eqnarray}
\lim_{\xi \rightarrow 0} {\cal H}(\alpha_c) & = & \frac{1}{2\sqrt{2\pi}}\left(\frac{
\beta^3}{\alpha_c}\right)^{1/4}\xi^{-1}\exp\left[-\frac{1}{2\xi^2}\left(
1-\sqrt{\beta\alpha_c}\right)^2\right]\mbox{,} \\
\lim_{\xi \rightarrow \infty} {\cal H}(\alpha_c) & = & \frac{z_o}{\sqrt{2\pi}}
\left(\frac{\beta}{\xi^2}\right)^{3/2}.
\end{eqnarray}
The distribution over $\alpha_c$ for different values of $\beta$ and $\xi$ 
is plotted in Figure 7. 
The behavior of ${\cal H}(\alpha_c)$ is analogous to that of $G(\alpha,e)$. In the 
limit of kicks much smaller than the relative velocity of the stars 
in the pre-SN orbit, conditions approximate the case of a symmetric explosion, 
and the distribution peaks at those values of $\alpha_c$ that are
consistent with such an explosion: $\alpha_c=1/\beta$ for $\beta \geq 0.5$ 
(Verbunt 1993), 
and $\alpha_c=2$ for $\beta < 0.5$. As the average kick velocity becomes
comparable to 
$V_r$, smaller orbital separations become more abundant, and for even 
larger kicks, 
a shrinkage  of the orbit relative to the pre-SN state is favored.  
 
\section{SYSTEMIC VELOCITIES}

During the supernova explosion,   
the post-SN system as a whole receives a velocity relative to 
the center of mass of the pre-SN binary. We derive the probability distribution
of the systemic velocities by performing a sequence of Jacobian transformations 
of the initial distribution of kick velocities. 

We choose to work in a reference frame, in which the three axes have the same 
orientation as the 
one shown in Figure 1, but which is centered on the center of mass of the 
system prior to the supernova explosion. In this frame the vector velocities 
of the two stars are: 
\begin{eqnarray}
\vec{V}_1 & = & (0,\frac{M_2}{M_1+M_2}V_r,0) \\
\vec{V}_2 & = & (0,-\frac{M_1}{M_1+M_2}V_r,0)\mbox{,} 
\end{eqnarray}
where $V_r$ is the magnitude of the relative velocity of the two stars. After 
the supernova explosion, $\vec{V}_2$ remains the same and $\vec{V}_1$ becomes:
\begin{equation}
\vec{V}_{NS}=(V_{kx},V_{ky}+\frac{M_2}{M_1+M_2}V_r,V_{kz}).
\end{equation}
Hence, the systemic velocity is:
\begin{eqnarray}
\vec{V}_{sys} & = & \frac{M_{NS}\vec{V}_{NS}+M_2\vec{V}_2}{M_{NS}+M_2} \nonumber \\
 & = & \frac{1}{M_{NS}+M_2}\left(M_{NS}V_{kx},M_{NS}V_{ky}-
\frac{(M_1-M_{NS})M_2}{M_1+M_2}V_r,M_{NS}V_{kz}\right).
\end{eqnarray} 
We define the dimensionless systemic velocity $v_{sys}\equiv V_{sys}/V_r$.  
Using equations (9), (10), and (11), we obtain:
\begin{equation}
v_{sys}^2 = \kappa_1 + \kappa_2~\frac{2\alpha -1}{\alpha}-\kappa_3 ~\cos \theta~\left[\alpha (1-e^2)
\right]^{1/2}\mbox{,}
\end{equation}
where:
\begin{eqnarray}
\kappa_1 & \equiv & \frac{M_1^2}{(M_1+M_2)^2}\mbox{,} \nonumber \\
\kappa_2  & \equiv & \frac{M_{NS}^2}{(M_{NS}+M_2)(M_1+M_2)}\mbox{,} \nonumber \\
\kappa_3  & = & 2\sqrt{\kappa_1~\kappa_2 }.
\end{eqnarray}

We derived above an expression (eq.\,[13]) describing the distribution 
of post-SN systems over orbital separation, $\alpha$, eccentricity, $e$, and 
orientation of the orbital plane, $\cos \theta$. Using equation (35) we can 
eliminate 
$\cos \theta$ and transform 
$g(\alpha,e,\cos \theta)$ into a distribution over $\alpha$, $e$, and the 
magnitude of the systemic velocity, $v_{sys}$:
\begin{eqnarray}
s(\alpha,e,v_{sys}) & = & \left(\frac{\beta}{2\pi \xi ^2}\right)^{3/2}~
\frac{4~e~v_{sys}}{\kappa_3 ~\alpha~(1-e^2)}\left[\left(\alpha-\frac{1}{1+e}\right)~
\left(\frac{1}{1-e}-\alpha\right)\right]^{-1/2} \nonumber \\
 &  & \times \exp\left[-\frac{1}{2\xi ^2}
\left(\beta\frac{2\alpha -1}{\alpha}+1\right)\right] 
\exp \left[\frac{\beta^{1/2}}{\kappa_3 \xi ^2}\left(\kappa_1+\kappa_2 \frac{2\alpha -1}
{\alpha}-v_{sys}^2
\right)\right] \nonumber \\
 &  & \times \left[1~-~\frac{\left(\kappa_1+\kappa_2 (2\alpha
-1)/\alpha-v_{sys}^2\right)^2}{\kappa_3 ^2\alpha (1-e^2)}\right]^{-1/2}.
\end{eqnarray}
The above expression is valid in the general case of $v_{kx}\neq 0$ and
$v_{kz}\neq 0$. The special case of $v_{kx}=0$ has already been
discussed and  corresponds to the pre-SN orbital separation becoming either
the periastron  or the apastron distance in the post-SN eccentric orbit (see
eq.\,[15]).  In the special case of $v_{kz}=0$, the plane of the orbit remains
unaffected  by the explosion, since the kick velocity is restricted in the
$x-y$ plane,  which is the orbital plane prior to the explosion 
(see eq.\,[11],\,[35] and Figure 1). 
The derivation of the probability density for 
these special cases is described in Appendix B.   

It is also interesting to study how the systemic velocity imparted to the 
binary during the supernova explosion correlates with the orbital separation 
after the circularization. 
We transform $s(\alpha,e,v_{sys})$ to 
a distribution over ($\alpha_c,\epsilon,v_{sys}$) (see eq.\,[20],\,[21],\,[22]):
\begin{eqnarray}
f(\alpha_c,\epsilon,v_{sys}) & = & 2\left(\frac{\beta}{2\pi \xi
^2}\right)^{3/2}~\exp\left[\frac{1}{2\xi
^2}\left(2\sqrt{\beta}\frac{\kappa_1+2\kappa_2 }{\kappa_3 }-(2\beta +1)\right)\right] \nonumber \\
 &  & \times v_{sys}~\left(2-\alpha_c-\epsilon\right)^{-1/2}~\exp\left[\frac{\beta}{2\xi ^2}
\left(\epsilon - 2\frac{\kappa_2 \epsilon+v_{sys}^2}{\kappa_3 \sqrt{\beta}}\right)\right] 
\nonumber \\
 &  & \times \left[\kappa_3 ^2\alpha_c - \left(\kappa_1+2\kappa_2 -\kappa_2 \epsilon-v_{sys}^2\right)^2
\right]^{-1/2},
\end{eqnarray}
and  we need to integrate over orbital energies, $\epsilon$. The above 
distribution has three poles in $\epsilon$; for clarity we rewrite it as:
\begin{eqnarray}
f(\alpha_c,\epsilon,v_{sys}) & = & \left(\frac{\beta}{2\pi \xi
^2}\right)^{3/2}~\frac{2}{\kappa_2 }~\exp\left[\frac{1}{2\xi
^2}\left(2\sqrt{\beta}\frac{\kappa_1+2\kappa_2 }{\kappa_3 }-(2\beta +1)\right)\right] \nonumber \\
 &  & \times v_{sys}~\exp\left[\frac{\beta}{2\xi ^2}
\left(\epsilon - 2\frac{\kappa_2 \epsilon+v_{sys}^2}{\kappa_3 \sqrt{\beta}}\right)\right]
\nonumber \\
 &  & \times (\epsilon - \lambda_1 )^{-1/2}~(\lambda_2 -\epsilon)^{-1/2}~(\lambda_3 -\epsilon)^{-1/2},
\end{eqnarray}
where
\begin{eqnarray}
\lambda_1  & = & \frac{\kappa_1+2\kappa_2 -v_{sys}^2-\kappa_3 \sqrt{\alpha_c}}{\kappa_2 }\mbox{,} \nonumber \\
\lambda_2  & = & \frac{\kappa_1+2\kappa_2 -v_{sys}^2+\kappa_3 \sqrt{\alpha_c}}{\kappa_2 }\mbox{,} 
\nonumber \\ 
\lambda_3  & = & 2-\alpha_c. \nonumber 
\end{eqnarray}
All three poles are numerically integrable except in the special case that:
\begin{equation}
\lambda_2 =\lambda_3  \Rightarrow v_{sys}=\sqrt{\kappa_1}+\sqrt{\alpha_c \kappa_2 }.
\end{equation}
In this case, the two-dimensional distribution $F(\alpha_c,v_{sys})$ becomes 
infinite along the line defined by equation (40). 
However, it is still integrable 
over $\alpha_c$ and $v_{sys}$, so that the final integral is finite. 
A sample distribution $F(\alpha_c,v_{sys})$ for a specific choice of $M_1$, 
$M_2$, 
and $\xi$, and for $c=0$ is shown in Figure 8. 
The spikes correspond to the pole (eq.\,[40]). 
The limits on $\alpha_c$ are in agreement with equation (27), 
while the limits on 
$v_{sys}$ are discussed in the next section. 

We can obtain the distribution of systemic velocities only, 
${\cal F}(v_{sys})$, 
by further 
integrating over $\alpha_c$. Sample distributions normalized to the total 
survival fraction for different values of 
$\xi$ are given in Figure 9. It is evident that the limits imposed 
on $v_{sys}$ are independent of the characteristics of the kick velocity 
distribution. 
As the magnitude of the 
kick velocity increases the systemic velocity remains restricted to a 
certain range of values specified by the stellar masses. 
Within this range, the distribution function of $v_{sys}$ shifts 
towards larger velocities 
as $\xi$ increases, reaching an asymptotic distribution  
for $\xi\gtrsim 3$, 
independent of $\xi$ (see Figure 9). As in the case of the one-dimensional 
distributions over eccentricities and orbital separations (see eq.\,[18] and 
[19]), this behavior is due to 
the fact that 
for $\xi \gg 1$, the exponential terms in equation (38)  
approach unity, and only the normalization constant depends on $\xi$.  
In this limit the normalized distribution of systemic velocities  
depends only on the 
stellar masses involved.  

\subsection{Limits on the systemic velocity}

Scrutiny of equation (38) shows that there exists an upper and a lower limit to 
the values of the systemic velocities, since the expression shown 
is real only if   
\begin{equation}
\kappa_3 ^2\alpha_c-(\kappa_1+2\kappa_2 -\kappa_2 \epsilon-v_{sys}^2)^2 > 0,
\end{equation}
or
\begin{equation}
(\kappa_1+2\kappa_2 -\kappa_2 \epsilon-\kappa_3 \sqrt{\alpha_c})^{1/2} < v_{sys} < 
(\kappa_1+2\kappa_2 -\kappa_2 \epsilon+\kappa_3 \sqrt{\alpha_c})^{1/2}. 
\end{equation}
In addition, we have already derived limits on $\epsilon$ and $\alpha_c$ 
(eq.\,[24] - [27]). 
By taking these into account, we can find the absolute lower and upper 
limits on $v_{sys}$:
\begin{eqnarray}
v_{sys} & < & \sqrt{\kappa_1}+\sqrt{2\kappa_2 } \nonumber \\ 
v_{sys} & > & \vert \sqrt{\kappa_1}-\sqrt{\alpha_c \kappa_2 }\vert > 
\sqrt{\kappa_1}-\sqrt{2\kappa_2 } \mbox{,} ~~~~~~~~ \mbox{if}~~ \frac{\kappa_1}{\kappa_2 } > 2. 
\end{eqnarray}
The last inequality is true if $M_1>2$\,$M_{NS}=2.8$\,M$_\odot$ (eq.\,[36]),
assuming a neutron star gravitational mass equal to $1.4$\,M$_\odot$. This 
condition is 
satisfied for all the progenitors of LMXBs 
forming via the He-star and direct supernova mechanisms (Kalogera \& Webbink\markcite{W96} 
1996; Kalogera\markcite{K96} 1996), and for most of the HMXB progenitors (Portegies-Zwart \& 
Verbunt 1996).  
Therefore, the maximum and minimum systemic
velocities are:
\begin{eqnarray}
v_{sys}^{max} & = & \frac{M_1}{M_1+M_2}+\frac{M_{NS}
\sqrt{2}}{(M_{NS}+M_2)^{1/2}(M_1+M_2)^{1/2}} \\
v_{sys}^{min} & = & \frac{M_1}{M_1+M_2}-\frac{M_{NS}
\sqrt{2}}{(M_{NS}+M_2)^{1/2}(M_1+M_2)^{1/2}}\mbox{,}~~~~~~~~\mbox{for}~~
M_1>2M_{NS} 
\end{eqnarray}
The above upper limit on $v_{sys}$ is in agreement with the one found by 
Brandt \& Podsiadlowski (1995), while the lower limit (eq.\,[45]) is stricter 
than theirs. Inspection of equation (44) shows that a maximum value of 
$v_{sys}^{max}$ exists in the limit that (i) $M_2$ is equal to zero and 
(ii) $M_1$ 
is equal to the minimum possible mass of a neutron star progenitor,  
$\sim 2.2$\,M$_\odot$ (e.g., Habets\markcite{H85} 1985). In this limit, 
the maximum of 
$v_{sys}^{max}$ is $\sim 2$, and hence the systemic velocity of a bound post-SN
binary can never exceed twice the value of the pre-SN relative orbital 
velocity, regardless of the magnitude of the kick velocity and 
the masses involved. 
Therefore, it becomes clear that high kick velocities do not 
necessarily result in high systemic velocities, as well.

Both upper and lower limits on systemic velocities can be understood 
physically. 
The maximum systemic 
velocity is acquired by that binary for which the neutron star receives 
a kick velocity oriented opposite to the pre-SN orbital velocity with  
a magnitude, such that its post-SN kinetic energy is just below its 
binding energy. 
The minimum systemic velocity is acquired by that binary in which the neutron 
star receives a kick velocity with the smallest possible magnitude needed to 
avoid disrupting  
the system due to mass loss. It is important to re-emphasize that neither  
the upper nor the lower limits depend on the kick 
velocity distribution. 
 
\section{SURVIVAL PROBABILITIES}

The total survival probability of a binary system 
with specific initial orbital characteristics is of interest to studies of 
the statistical properties of an entire population of binaries.  
We can obtain this survival probability by integrating over the distribution 
of circularized dimensionless orbital separations, ${\cal H}(\alpha_c)$. This
integration 
can only be performed numerically, but it is straightforward  since
the function has no poles and the limits of integration are well-defined 
(eq.\,[27]). 

Clearly, kick velocities to neutron stars may 
bind systems that would otherwise be disrupted, or disrupt those 
that would have remained bound. If the average kick velocity is large 
compared to the initial
relative orbital motion of the binary components then survival depends on the
small probability that the kick velocity is itself small and directed opposite
to the original motion of the collapsing component. If the ratio of kick
velocity to initial relative orbital velocity is
small, then the survival
rate of systems that would otherwise be 
disrupted falls very rapidly as this
ratio decreases. Asymptotically, we have, respectively 
\begin{eqnarray}
\lim_{\xi \rightarrow \infty} \int_{2c/(1+c)}^2 H(\alpha_c )d\alpha_c  & = &
\frac{4}{3\sqrt{\pi}}~\frac{(1-c)^{\slantfrac{3}{2}}}{(1-c^2)}~
\left(\frac{\beta}{\xi^2}\right)^{\slantfrac{3}{2}} \nonumber \\
\lim_{\xi \rightarrow 0} \int_{2c/(1+c)}^2 H(\alpha_c )d\alpha_c  & = &
\frac{1}{2}\left[1~+~\mbox{erf} \left(\frac{\beta - \frac{1}{2}}{\xi\sqrt{2}}\right)\right]
\end{eqnarray}
 
The survival fractions for a wide range of values of $\xi$, and for two 
different values of $\beta$, are shown of Figure 10. In this illustration, 
$\xi$ is varied by keeping the r.m.s. kick 
velocity constant, while allowing the pre-SN orbital separation 
to vary. Among those systems that
would remain bound if collapse were symmetric ($M_1=3.8$\,M$_\odot$ and 
$M_2=1.0$\,M$_\odot$), kick
velocities will tend to unbind widely separated binaries, for which the
relative orbital motion falls below the kick velocity. 
Among systems that suffer so much mass loss in a supernova
that they would otherwise be disrupted ($M_1=8.6$\,M$_\odot$ and 
$M_2=1.0$\,M$_\odot$), 
kick velocities will favor the
survival of binary systems in just that range of separations where the
relative orbital velocity is comparable to the mean kick velocity. 

\section{CONCLUSION}

The expressions derived here provide a tool necessary
in analytical population syntheses of neutron star binaries.  The additional
step needed in such syntheses is to convolve the distribution over post-SN parameters with the
distribution of pre-SN binaries over masses and orbital separations (see also
Wijers et al. 1992). This link depends on the type of final  systems and the
specifics of their formation mechanism. In addition, the  distribution
functions of systemic velocities and their correlation with  orbital
separations and eccentricities (or circularized orbital separations)  can be
used in studying the motion of neutron star binaries in the Galactic 
potential, and in modeling their spatial distribution in the Galaxy. 

The results of the study presented in this paper have a number of important 
implications concerning the population of neutron star binaries: 

There exists a correlation between orbital
separations and eccentricities, which is independent of the 
characteristics of the
binary or the magnitude of the kick velocity.
For post-SN
orbits much wider than the pre-SN orbit, the total energy of the binary
significantly increases, and the system remains bound only
in a highly eccentric orbit. On the other hand, the eccentricity may be low
($e\lesssim 0.4$) only if the post-SN orbital separation is comparable to
that before the explosion. 
The discovery of a double neutron star system of modest eccentricity could 
therefore be 
used to infer  
the size of the orbit of its progenitor, provided that the gravitational 
radiation decay time scale for the
orbit were long enough for such losses to be negligible.

The ratio of the post-SN systemic velocity,
$V_{sys}$, to the pre-SN relative orbital velocity, $V_r$, is restricted in 
a relatively narrow range of values. Both lower and upper limits depend
only on 
the stellar masses involved. For the ranges of
progenitor masses relevant to HMXBs, LMXBs, and double neutron star binaries
we find 
$V_{sys}^{max}\lesssim 1.5\,V_r$. 
Since LMXB progenitors are more
tightly bound than those of HMXBs, and hence have higher relative orbital velocities than HMXB
progenitors, the systemic velocities of LMXBs are
expected to be higher than those of HMXBs. 
It is also clear that measurements 
of systemic velocities of neutron
star binaries do not necessarily reveal information about the kick velocities 
imparted to neutron stars in individual systems. 
Instead, they can be
used to infer typical relative orbital velocities prior to the
supernova explosion. 

Although the allowed range of systemic velocities is independent of the kick 
velocity, the probability distribution within this range does depend on the 
r.m.s. magnitude of kick velocities relative to the pre-SN orbital 
velocities. In the limit of very small kicks the distribution sharply peaks 
at values close to the lower end of the range. As the r.m.s. of the kick 
velocities increases 
the distribution becomes broader and its peak shifts to higher velocities. For 
kicks much higher than the pre-SN relative orbital velocities, the shape of the 
distribution 
remains unaffected, and further increases of the r.m.s. kick velocity 
only decrease the 
binary survival rate, without altering the velocity distribution of bound
post-SN  
systems. 
Measurements of the systemic velocities of neutron star binaries will 
possibly prove quite significant in distinguishing between symmetric and 
asymmetric explosions with high kicks imparted to the neutron stars.  

The incidence of X-ray binaries in globular clusters relates to their smallest
possible systemic velocity and to how this velocity compares with the escape
velocity from the cluster. For LMXBs formed via the explosion of the He-star
remnant of a common envelope phase, 
typical parameters for the progenitors yield $V_{sys}^{min} \simeq 
100\,$km s$^{-1}$
(Kalogera \& Webbink 1996).
The direct-SN channel (Kalogera\markcite{K196} 1996; Kalogera 1996) is fed by 
binaries with orbits which are much wider, but still small enough to 
avoid disruption by 
dynamical interactions. 
Typical parameters in this case yield $V_{sys}^{min} \simeq 20\,$km s$^{-1}$. 
Estimates of the escape velocities from the cores of globular clusters 
that
contain LMXBs range from $30$\,km\,s$^{-1}$ to $60$\,km\,s$^{-1}$ (for
NGC 1851, 6440, 6441, 6624, M15, and Lil 1); more loosely bound clusters
such as Ter 1 and 2 have central escape velocities of the order of
$10$\,km\,s$^{-1}$ (Webbink\markcite{W85} 1985; van Paradijs\markcite{V95} 1995).
It is therefore clear that post-SN binaries formed in globular clusters
from primordial binaries via the He-SN channel 
have a very small chance of remaining in the
clusters and becoming X-ray binaries. 
LMXBs formed via the direct-SN channel, on the other hand, 
will remain in the clusters, but their
formation rate is too low to account for a significant fraction of the LMXB 
population in globular clusters. 
Barring accretion-induced collapse as an alternative formation channel, 
it therefore appears that 
low-mass X-ray binaries 
observed in globular clusters must have formed through  
stellar exchanges and
captures, rather than directly from primordial binaries.
 
We have already applied the analytical method presented here to study 
low-mass X-ray binaries formed via different evolutionary channels. The 
results of these population synthesis calculations will be presented elsewhere 
(Kalogera \& Webbink 1996; Kalogera 1996). 

\acknowledgments

I am grateful to Ron Webbink for initiating this project, for many helpful 
discussions during the course of this study, and for carefully reading the 
manuscript. I would also like to thank Dimitrios Psaltis and Fred Lamb 
for valuable 
comments and discussions.  
This work was supported by National Science Foundation under
grant AST92-18074.

\appendix

\begin{center}
\large
{\bf Appendices}
\normalsize 
\end{center}

\section{Notation}

The symbols of the most important physical parameters are:

\noindent
$M_1$: mass of the exploding star,

\noindent
$M_2$: mass of the companion to the exploding star and to the 
neutron star,

\noindent
$M_{NS}$: mass of the neutron star,

\noindent
$R_2$: radius of the companion to the neutron star,

\noindent
$\alpha $: ratio of the post-SN orbital separation to the pre-SN 
separation,

\noindent
$e$: eccentricity of the post-SN orbit,

\noindent
$\theta $: angle between the pre-SN and post-SN orbital planes,

\noindent
$\beta$: ratio of the total mass after the explosion to that before,

\noindent
$\xi $: ratio of the standard deviation of the kick velocity 
distribution to the relative orbital velocity before the explosion, 

\noindent
$c$: ratio of the radius of the companion star to the pre-SN 
orbital separation,

\noindent
$\alpha_c$: ratio of the circularized orbital separation to  
that before the explosion,

\noindent
$\epsilon $: ratio of the post-SN binding energy to that of a system 
consisting of the neutron star and the companion star in an orbit with the 
pre-SN orbital 
separation, 

\noindent
$v_{sys}$: ratio of the systemic post-SN velocity to the pre-SN relative 
orbital velocity.  

The symbols of the various distribution functions over dimensionless 
orbital parameters, and respective equations, in which the corresponding 
expressions are given, are: 

\noindent
$g(\alpha,e,\cos \theta)$: distribution function of post-SN orbital 
separations, eccentricities, and cosine of the angles between pre- and post-SN 
orbital planes; equation (13),

\noindent
$G(\alpha,e)$: distribution function of post-SN orbital separations and
eccentricities; equation (14),

\noindent
${\cal J}(e)$: distribution function of post-SN eccentricities for 
$\xi \gg 1$; equation (18),

\noindent 
${\cal G}(\alpha)$: distribution function of post-SN orbital separations for 
$\xi \gg 1$; equation (19),

\noindent
$H(\alpha_c,\epsilon)$: distribution function of circularized orbital 
separations and post-SN binding energies; equation (23),

\noindent
${\cal H}(\alpha_c)$: distribution function of circularized orbital 
separations; equation (28),

\noindent
$s(\alpha,e,v_{sys})$: distribution function of post-SN orbital separations, 
eccentricities, and systemic velocities; equation (37),

\noindent 
$f(\alpha_c,\epsilon,v_{sys})$: distribution function of circularized 
orbital separations, post-SN binding energies, and systemic velocities; 
equation (38). 

\section{Effect of an impulse velocity}

According to the  numerical calculations performed by Fryxell \& Arnett (1981), 
approximately half of the momentum carried by the ejecta intersecting the 
companion is transferred to it. If $E_{SN}$ is the kinetic energy of the 
ejecta and $V_{imp}$ is the velocity imparted to the companion of mass $M_2$, 
then:
\begin{equation}
M_2V_{imp} \simeq \frac{1}{2} (M_{1}-M_{NS})\left(\frac{2E_{SN}}{(M_{1}-M_
{NS})}\right)^{\frac{1}{2}}\left(\frac{\pi R_2^2}{4\pi A_{i}^2}\right),
\end{equation}
where $R_2$ is the radius of the companion.
 
We assume that the impulse velocity, $V_{imp}$, is given to the companion in 
a direction along the line connecting the two stars, pointing away from the 
neutron star. In the reference frame shown in Figure 1,  
the velocity of the neutron star relative to 
that of the companion immediately after the explosion is then:
$\vec{V}'=(V_{kx}+V_{imp},V_{ky}+V_r,V_{kz})$. 
Following the same procedure as that described in \S\,2, we calculate the 
distribution of circularized orbital separations:
\begin{equation}
{\cal H}_{imp}(\alpha_c)=\frac{\beta \pi}{(2\pi\xi^2)^{3/2}}~\exp \left(-\frac{\beta 
\alpha_c +1}{2\xi^2}\right)~I_o\left(\frac{\sqrt{\beta\alpha_c}}{\xi^2}\right)~
S_1,
\end{equation}
where
\begin{eqnarray}
S_1 & \equiv & \int_{z_-}^{z_+}\exp \left(-\frac{z^2}{2\xi^2}\right)~dz~=~
\xi\sqrt{\frac{\pi}{2}}\left[\mbox{erf}\left(\frac{z^+}{\xi\sqrt{2}}\right)-
\mbox{erf}\left(\frac{z^-}{\xi\sqrt{2}}\right)\right], \nonumber \\
z_{\pm} & = & v_{imp}\pm \sqrt{\beta\left(2-\alpha_c-\frac{2c-\alpha_c}
{c^2}\right)},~~~~~~~~
\mbox{if}~~ \frac{2c}{1+c}<\alpha_c<2c \nonumber \\
 & = & v_{imp} \pm \sqrt{\beta (2-\alpha_c)},~~~~~~~~~~~~~~~~
\mbox{if}~~2c<\alpha_c<2 
\end{eqnarray}
and $v_{imp}\equiv V_{imp}/V_r$, $V_r$ being the relative orbital velocity 
of the pre-SN binary. 

Integrating the above distribution over $\alpha_c$, we find the survival 
probability when the impulse velocity is taken into account. In Figure 11
we plot the survival probability as a function of the pre-SN 
orbital separation, 
$A_i$, with and without the effect of the impulse velocity, 
for the specific choice of $M_1=4$\,M$_\odot$, $M_2=1$\,M$_\odot$ 
(typical of an LMXB progenitor), 
and $<V_k^2>^{1/2}=450$\,km\,s$^{-1}$. It is clear that  
the survival fraction decreases due to the impulse being imparted to the 
companion only when the orbital separation is very small, 
$A_i\lesssim 3$\,R$_\odot$. This separation is smaller than typical values of 
the pre-SN orbital separations of progenitors of X-ray binaries or 
double neutron star systems. 

\section{Distribution over eccentricities and orbital separations 
for special cases}

\subsection{$V_{kx}=0$}

Following the same procedure as in the general
case described in \S\,2.1, we transform the two dimensional 
distribution of kicks ${
p}_{yz}(v_{ky},v_{kz})$ into
$g'(e,\cos \theta)$, and then integrate over $\cos \theta$.
The eccentricity is
given by:
\begin{equation}
e~=~\pm \left[\frac{(v_{ky}+1)^2+v_{kz}^2}{\beta}-1\right]\mbox{,}
\end{equation}
where the plus sign corresponds to $\alpha (1-e)=1$, the minus sign to
$\alpha (1+e)=1$, and $\cos \theta$ is still given by equation (11).
The form of the integral of $g'(e,\cos \theta)$ over $\cos \theta$ 
is the same as
in the  general case, so we obtain:
\begin{equation}
{\cal J}'(e)~=~{\cal J}'_{+}(e)+{\cal J}'_{-}(e)\mbox{,}
\end{equation}
where:
\begin{eqnarray}
{\cal J}'_{\pm}(e) & = & \frac{\beta}{2\xi ^2}~\exp \left[-\frac{1}{2\xi^2}\left(1+
\beta(1\pm e)\right)\right]~I_o(z_e)\mbox{,} \nonumber \\
z_e & \equiv & \frac{\left[\beta (1\pm e)\right]^{1/2}}{\xi^2}\mbox{.}
\end{eqnarray}
The plus sign again corresponds to $A_i$ being the periastron distance, 
and the minus sign to $A_i$ being the apastron distance, in the post-SN
orbit. The corresponding distribution over $\alpha$ is:
\begin{eqnarray}
{\cal G}'(\alpha) & = & \frac{1}{2\xi^2}~\frac{\beta}{\alpha^2}~\exp \left[-\frac{1}
{2\xi^2}\left(1+\beta\frac{2\alpha -1}{\alpha}\right)\right]~I_o(z_\alpha)
\mbox{,} \nonumber \\
z_\alpha & \equiv & \frac{[\beta (2\alpha -1))/\alpha]^{1/2}}{\xi^2}\mbox{.}
\end{eqnarray}

\subsection{$V_{kz}=0$}

In this case it is $\cos \theta =\pm 1$ (eq.\,[11]), that is, the post-SN 
orbital 
plane is either the same as the pre-SN one ($\theta =0$), 
or it has been rotated by 
an angle $\theta =\pi$, and the stars orbit in a retrograde sense after the 
explosion. 
We transform the two-dimensional distribution of kicks 
$p_{xy}(v_{kx},v_{ky})$ into a distribution of dimensionless orbital 
separations and systemic velocities, $s'(\alpha,v_{sys})$. Using 
equations (9), (10), and (35) we obtain the expressions 
relating the four variables: 
\begin{eqnarray}
v_{kx}^2 & = & \beta\frac{2\alpha-1}{\alpha}-\frac{\beta}{\kappa_3 ^2}
\left(v_{sys}^2-\kappa_1-\kappa_2 \frac{2\alpha-1}{\alpha}\right)^2 \\
v_{ky}^2 & = & \left[1\pm \frac{\sqrt{\beta}}{\kappa_3 }\left(v_{sys}^2-\kappa_1-\kappa_2 \frac
{2\alpha-1}{\alpha}\right)\right]^2,
\end{eqnarray}
where the plus sign corresponds to $\cos \theta =-1$ and the minus sign to 
$\cos \theta =1$. 
We calculate the necessary Jacobian and find:
\begin{equation}
s'(\alpha,v_{sys}) = s'_+(\alpha,v_{sys}) + s'_-(\alpha,v_{sys}),
\end{equation}
where: 
\begin{eqnarray}
s'_{\pm}(\alpha,v_{sys}) & = & \frac{2\sqrt{\beta}}{\kappa_3 \pi\xi^2}~\frac{v_{sys}}
{\alpha^2}~\left[\frac{2\alpha -1}{\alpha}-\frac{\left(v_{sys}^2-\kappa_1-\kappa_2 \frac{2\alpha-1}{\alpha}\right)^2}{\kappa_3 ^2}\right]^{-1} \nonumber \\
 &  & \times \exp\left[-\frac{1}{2\xi^2}\left(1\pm\frac{\sqrt{\beta}}{\kappa_3 }\left(v_{sys}^2-\kappa_1-\kappa_2 \frac{2\alpha-1}{\alpha}\right)\right)^2\right] \nonumber \\
 &  & \times \exp\left[-\frac{\beta}{2\xi^2}\left(\frac{2\alpha-1}{\alpha}-
\frac{\left(v_{sys}^2-\kappa_1-\kappa_2 \frac{2\alpha-1}{\alpha}\right)^2}{\kappa_3 ^2}\right)
\right].
\end{eqnarray}

\newpage

\begin{center}
\large
{\bf Figure Captions}
\normalsize
\end{center}

\noindent
Fig.\,1--- Geometry of the binary system and reference frame adopted in the 
calculations. The orbital plane of the pre-SN binary coincides with the plane 
of the page, that is the $x$-$y$ plane. 

\noindent
Fig.\,2--- Limits on the parameter space $\alpha - e$ of post-SN systems
for a range of values of $c$, the ratio of the radius of the companion to 
the orbital separation prior to the explosion. For point stars ($c=0$), the 
allowed 
parameter space is restricted between the two thick lines, which correspond to 
limits due to the
geometrical constraint $1/(1-e)>\alpha>1/(1+e)$. The thin lines, which 
correspond to the  
constraint barring a physical collision, 
$\alpha>c/(1-e)$, impose a second, more stringent, lower 
limit to $\alpha$.

\noindent
Fig.\,3--- Distribution of post-SN systems over dimensionless orbital
separations $\alpha$ and eccentricities $e$, for $\beta=0.6$ and $\xi=1.0$.

\noindent
Fig.\,4--- Distribution of post-SN systems over eccentricities $e$ for systems
that (a)  would remain bound ($\beta=0.6$), or (b) be disrupted ($\beta=0.4$), 
in the  case of a symmetric explosion,
for different values of
$\xi$.  The probability density  for $\xi=10^{-3}$ in (a) has been reduced by a
factor of 
$100$. 

\noindent
Fig.\,5--- Distribution of post-SN systems over dimensionless orbital
separations for systems that (a) would remain bound ($\beta=0.6$), 
or (b) be disrupted ($\beta=0.4$), for different 
values of
$\xi$.

\noindent
Fig.\,6--- Distribution of post-SN binaries over (a) eccentricities, for 
$\beta=0.6$ and $\xi=1.0$, and over (b) dimensionless orbital separations, 
for $\beta=0.6$ and $\xi=1.0$, for different values of $c$.

\noindent 
Fig.\,7--- Distribution of post-SN binaries over dimensionless 
circularized orbital separations, $\alpha_c$, 
for different values of $\beta$ and $\xi$.

\noindent
Fig.\,8--- Distribution of post-SN systems over dimensionless 
orbital separations of 
circularized orbits, $\alpha_c$, and systemic velocities, $v_{sys}$, 
for $\beta=0.6$ and $\xi=1$. 

\noindent
Fig.\,9--- Distribution of post-SN systems over dimensionless systemic 
velocities, $v_{sys}$, for $M_1=3$\,M$_\odot$, $M_2=1$\,M$_\odot$ 
($\beta=0.6$), 
and different values of $\xi$. 

\noindent
Fig.\,10--- Fraction of systems that survive the supernova event, at which kick
velocities with $<V_k^2>^{1/2}=100$\,km\,s$^{-1}$ are imparted to the neutron
star, as a function of pre-SN orbital separation for a $1$\,M$_\odot$
secondary and two different masses of the exploding star. In the case of a 
symmetric explosion, systems with $M_1=3.8$\,M$_\odot$ 
remain marginally bound, while
systems with $M_1=8.6$\,M$_\odot$ do not survive.

\noindent
Fig.\,11--- Fraction of systems that survive the supernova event
as a function of pre-SN orbital separation with ({\it dotted line}) and
without ({\it solid line}) an impulse velocity imparted to the secondary. 

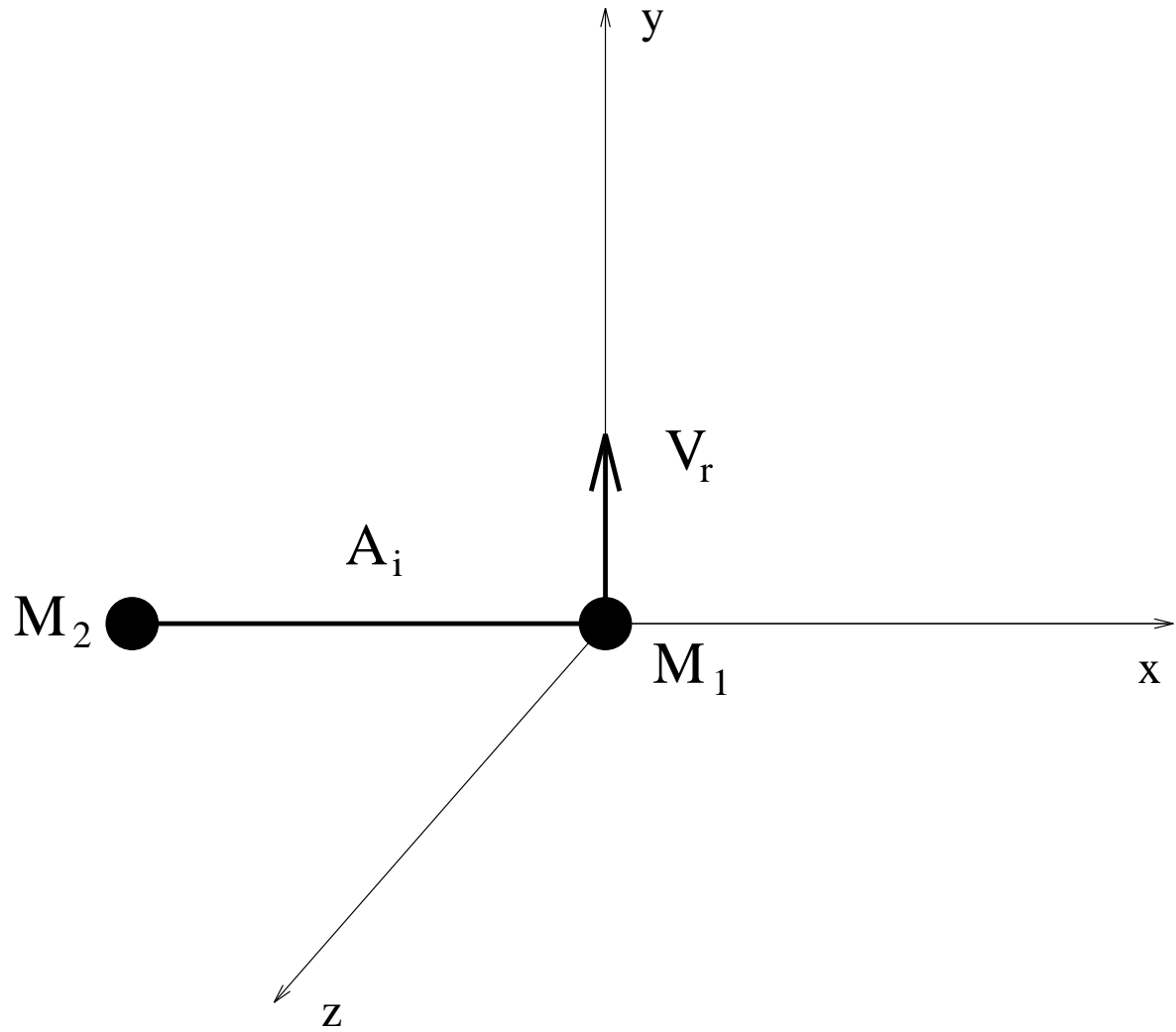
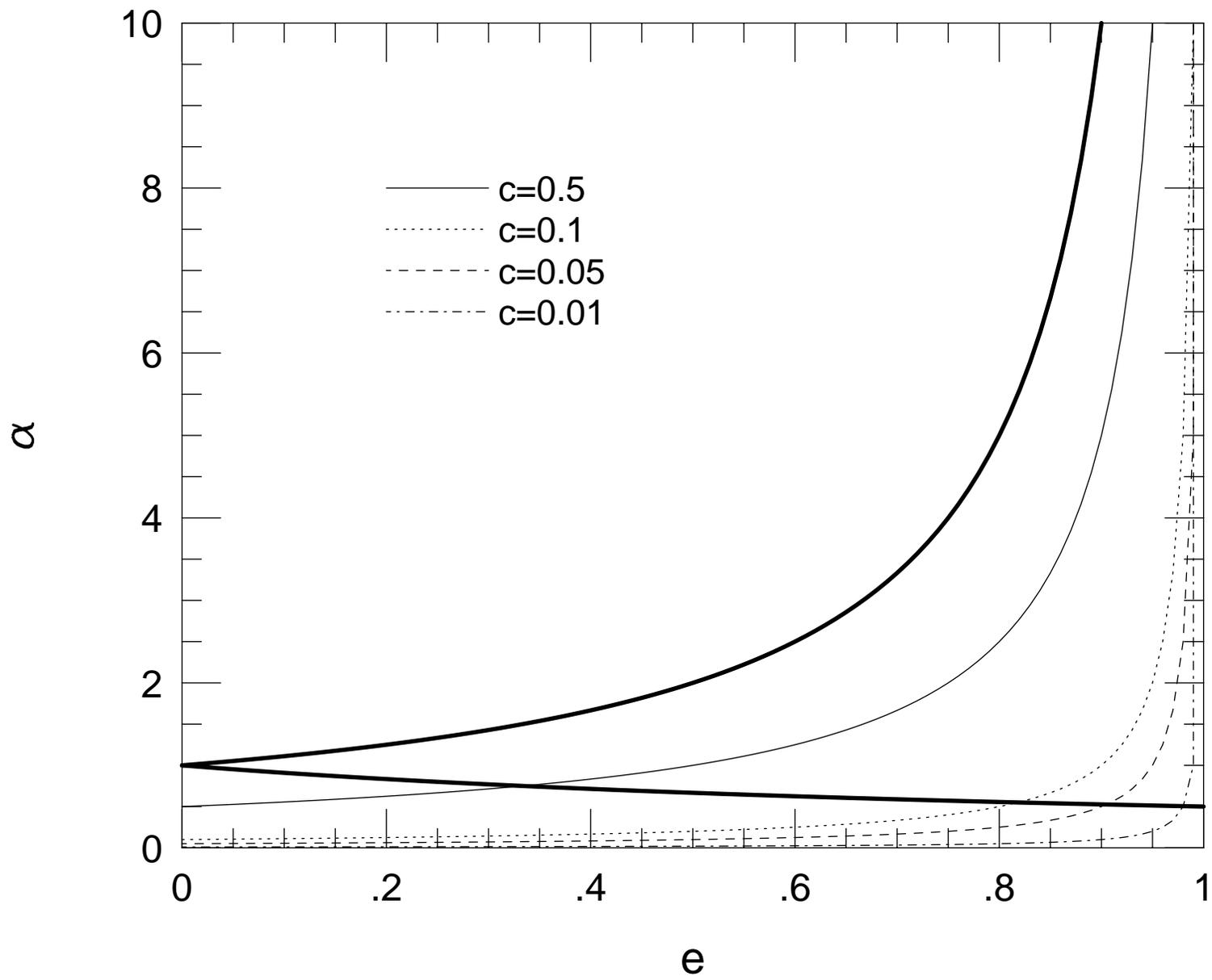
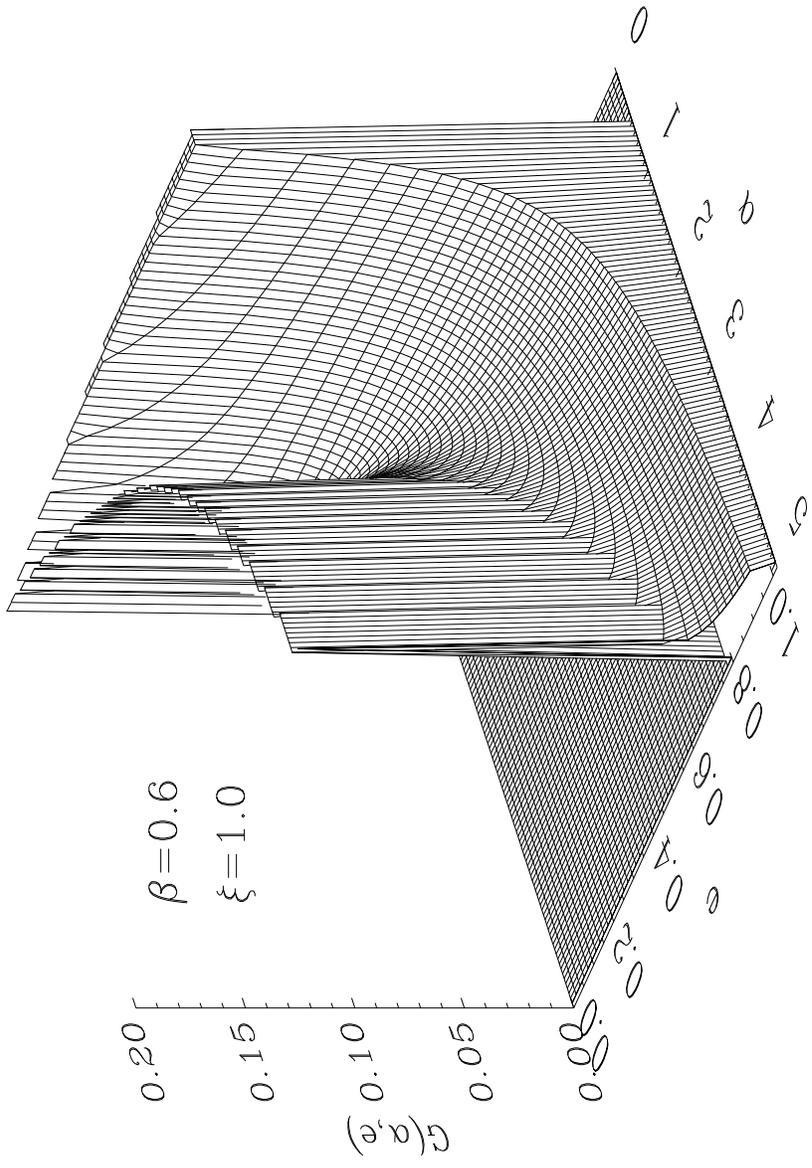
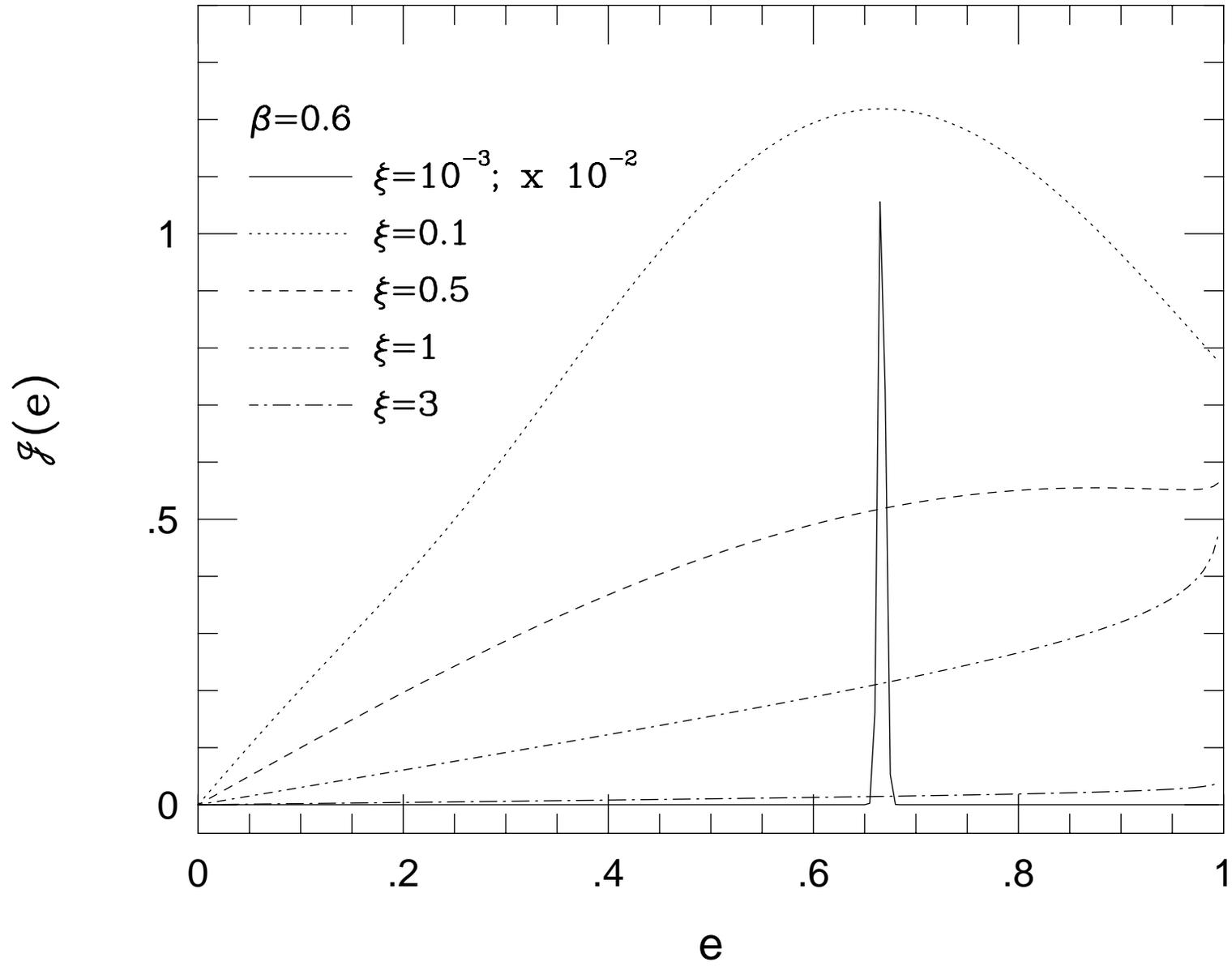
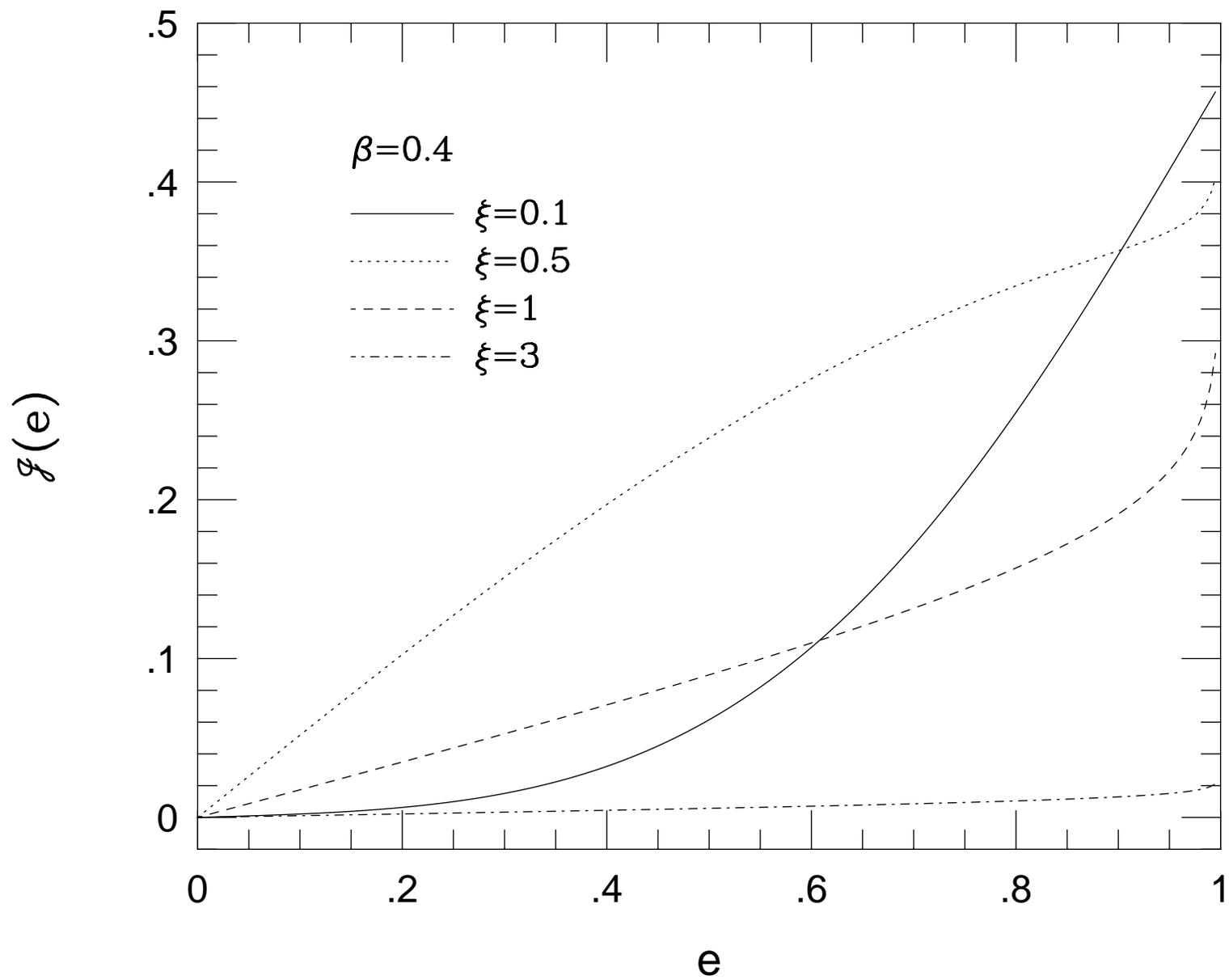
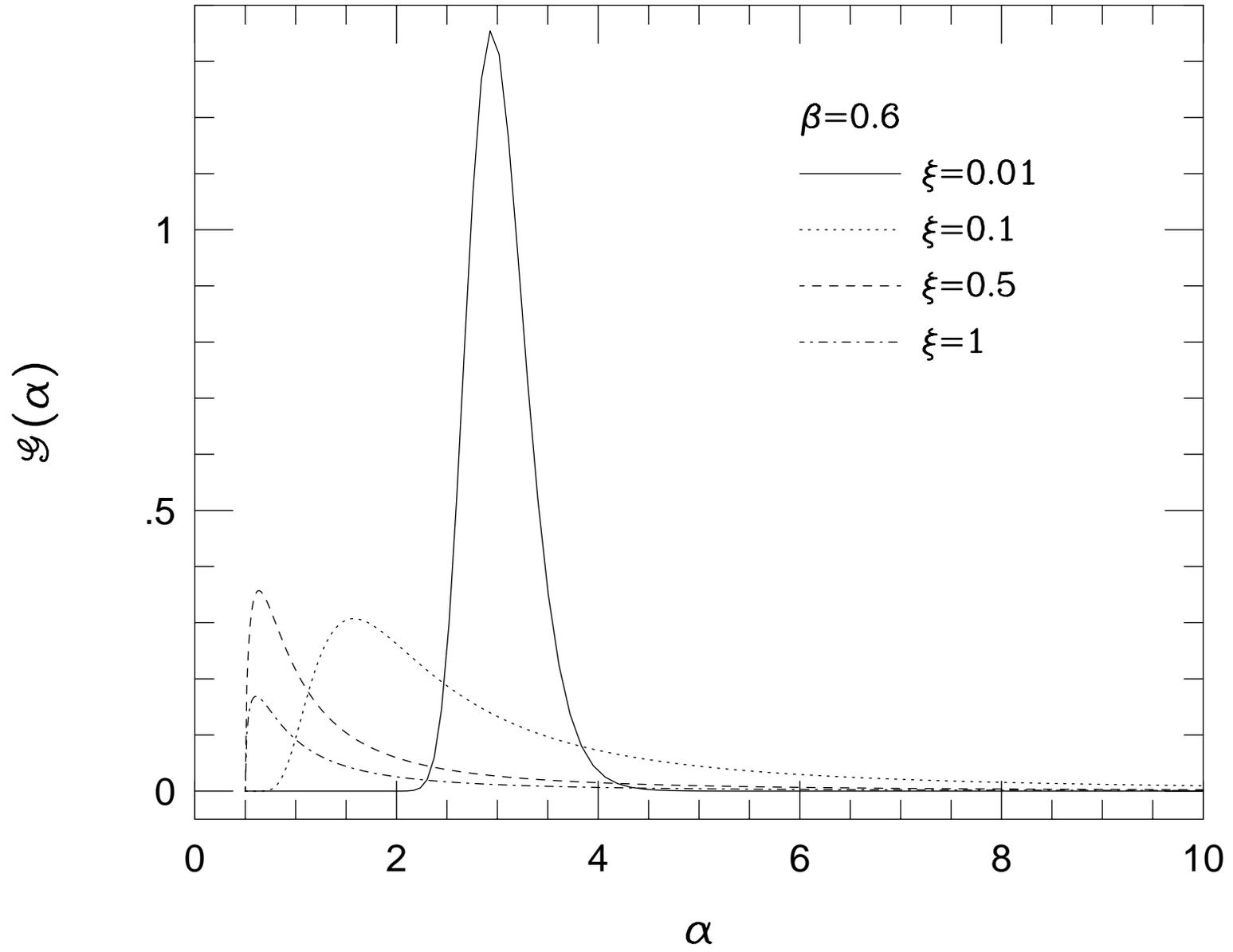
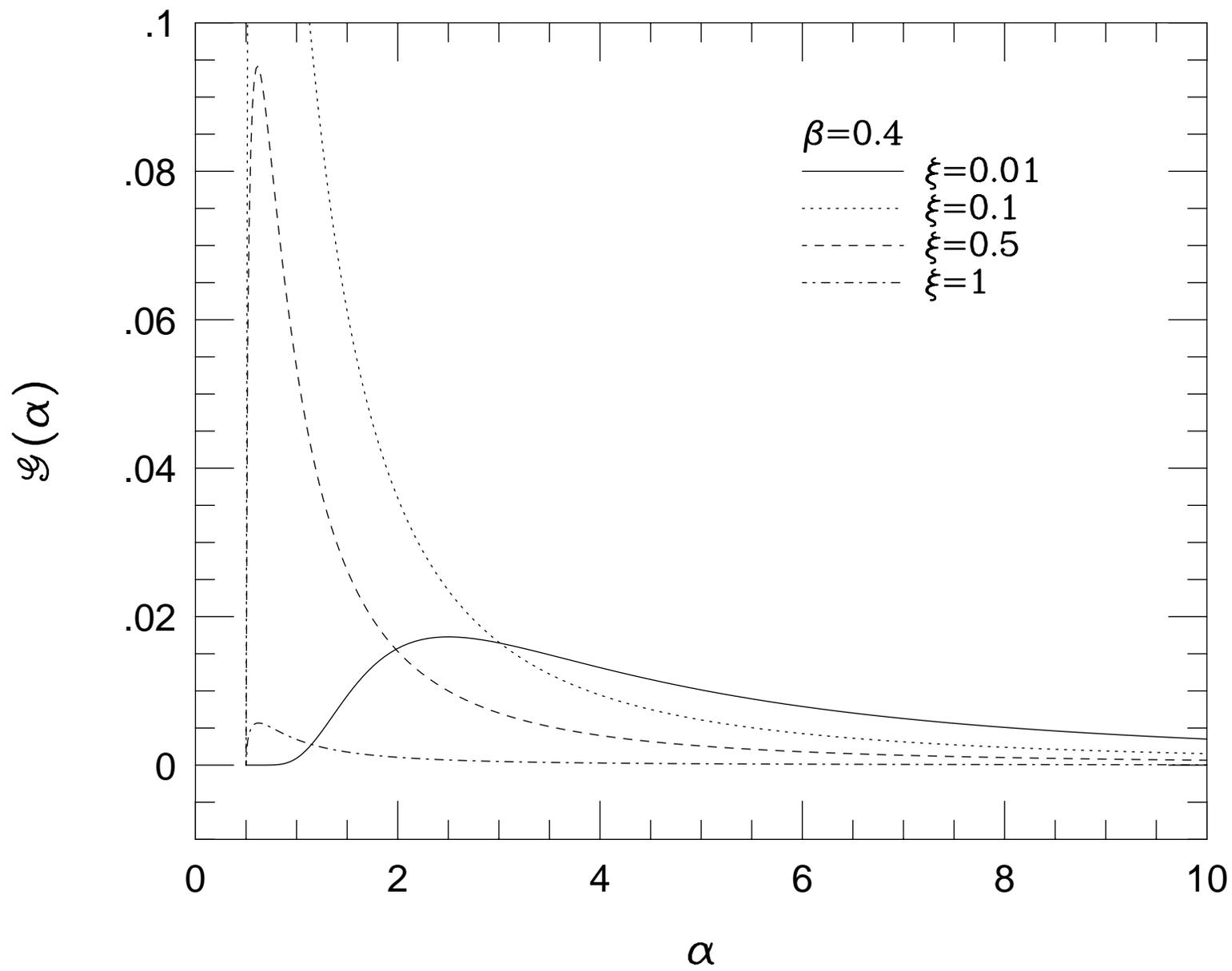
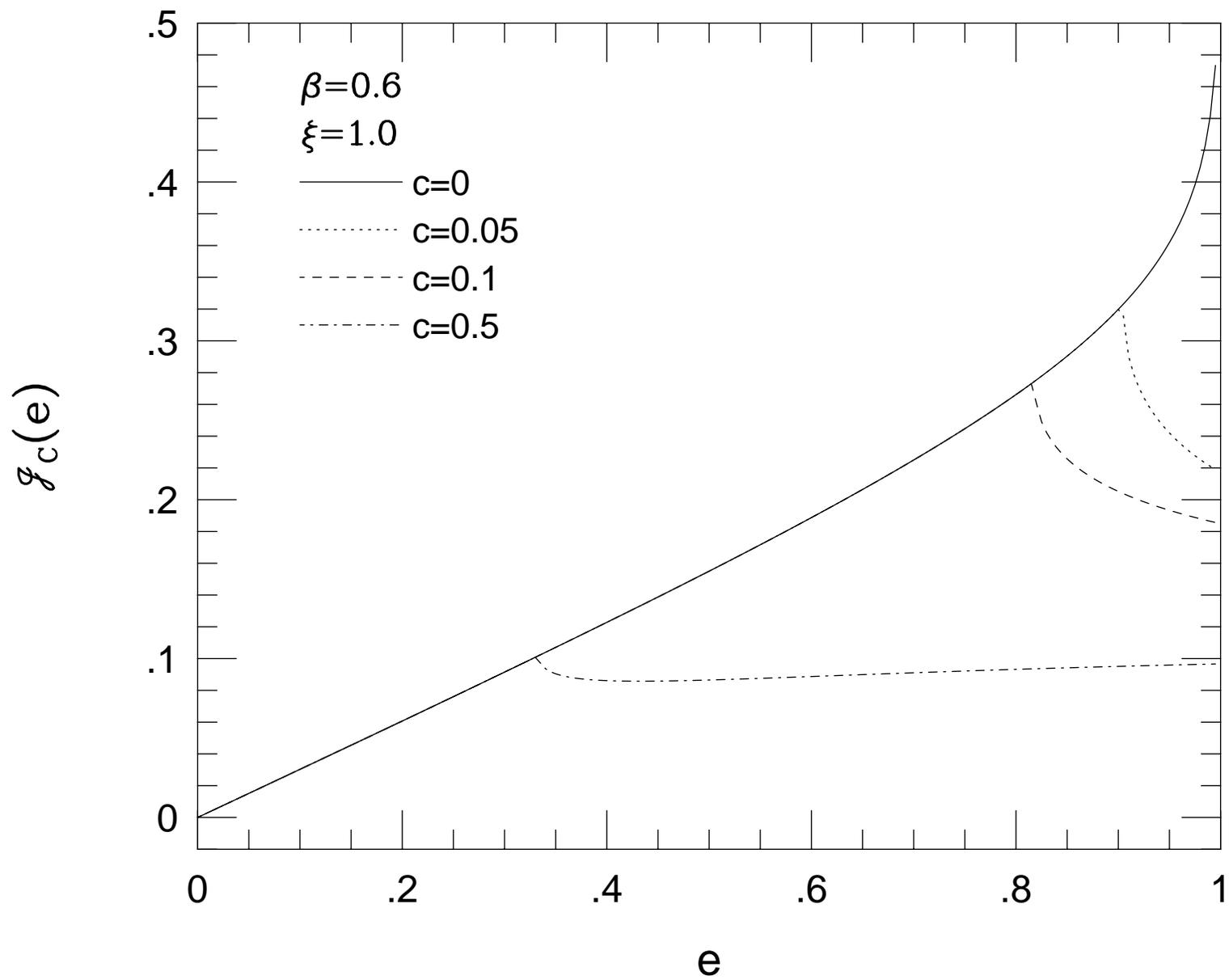
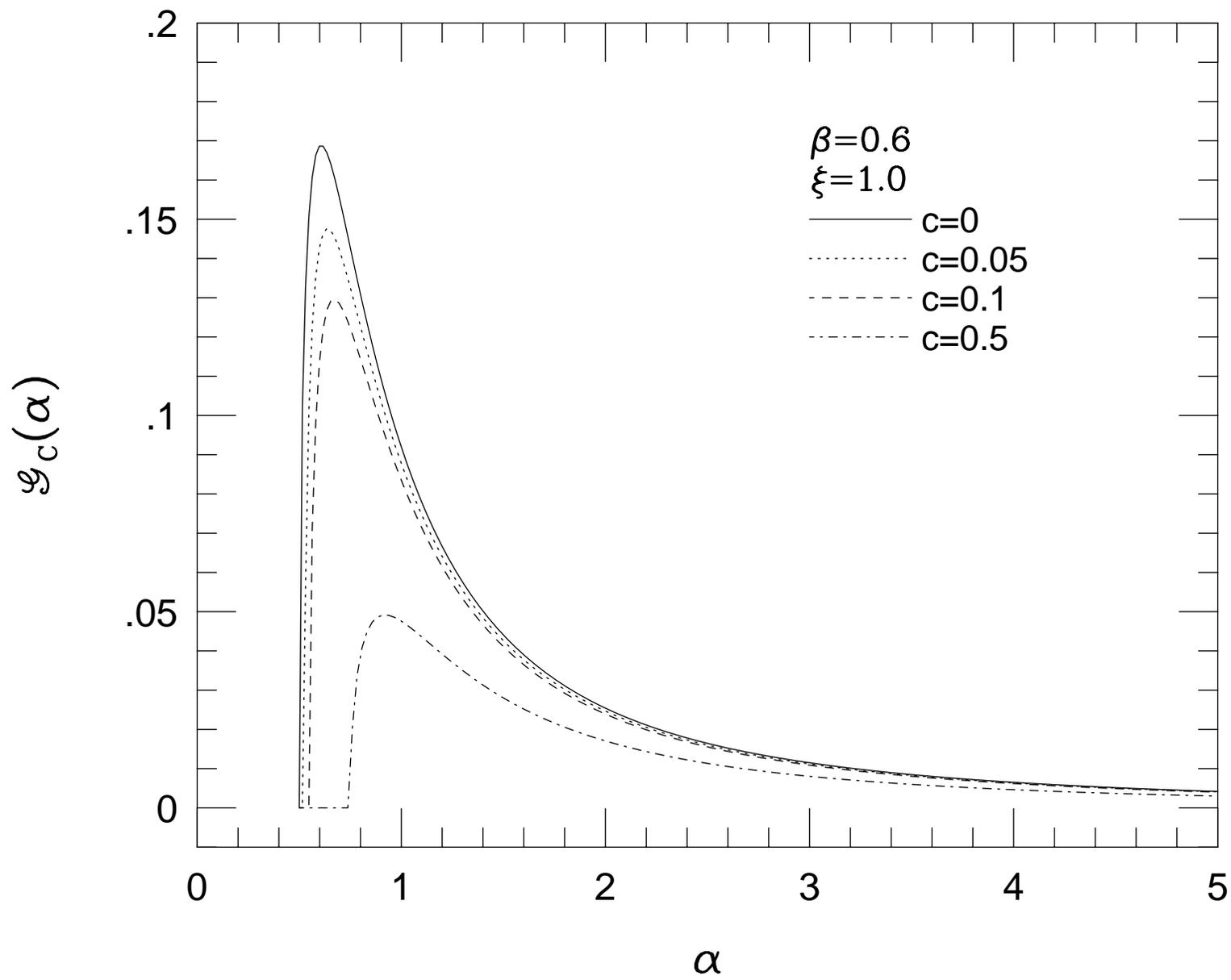
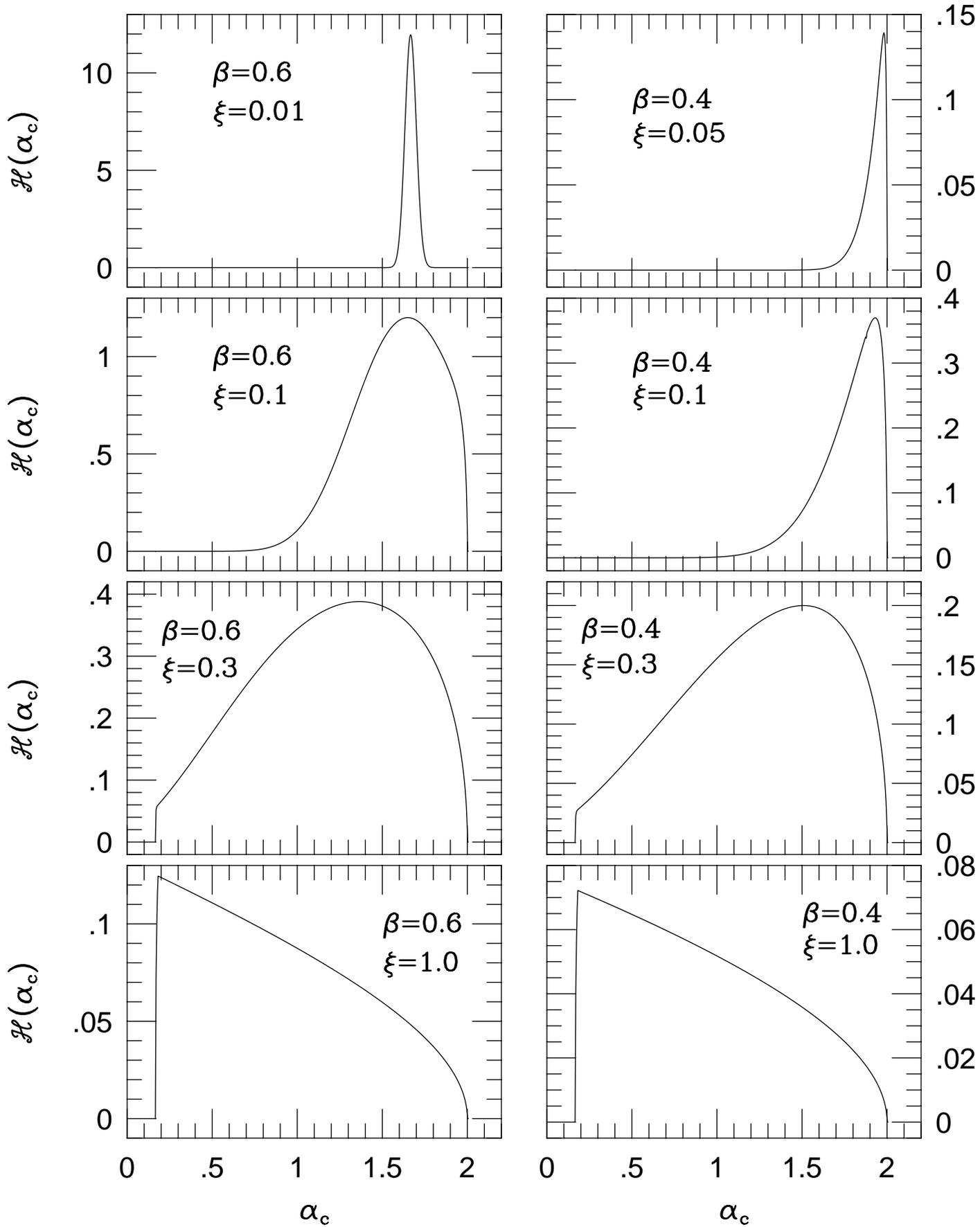
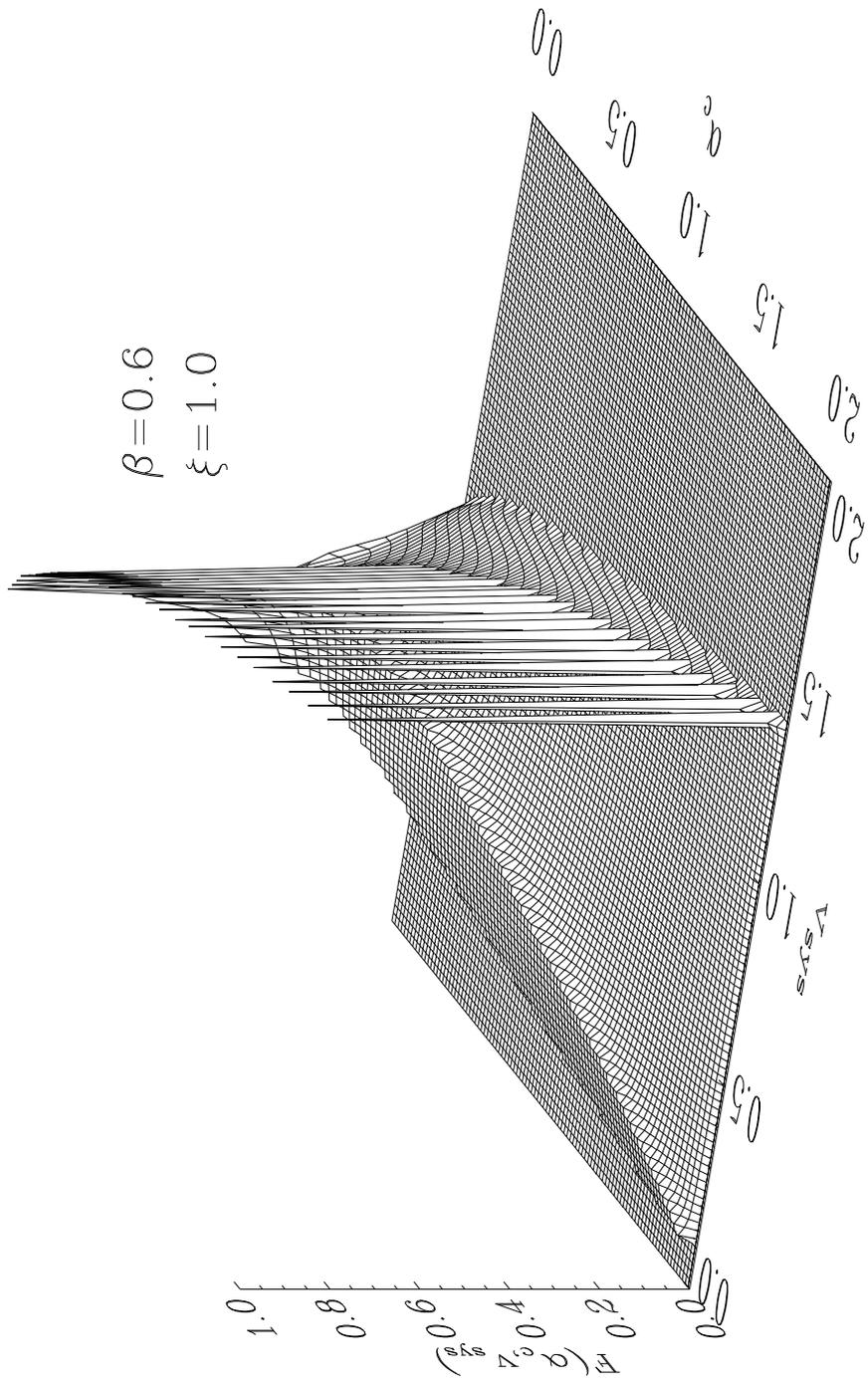
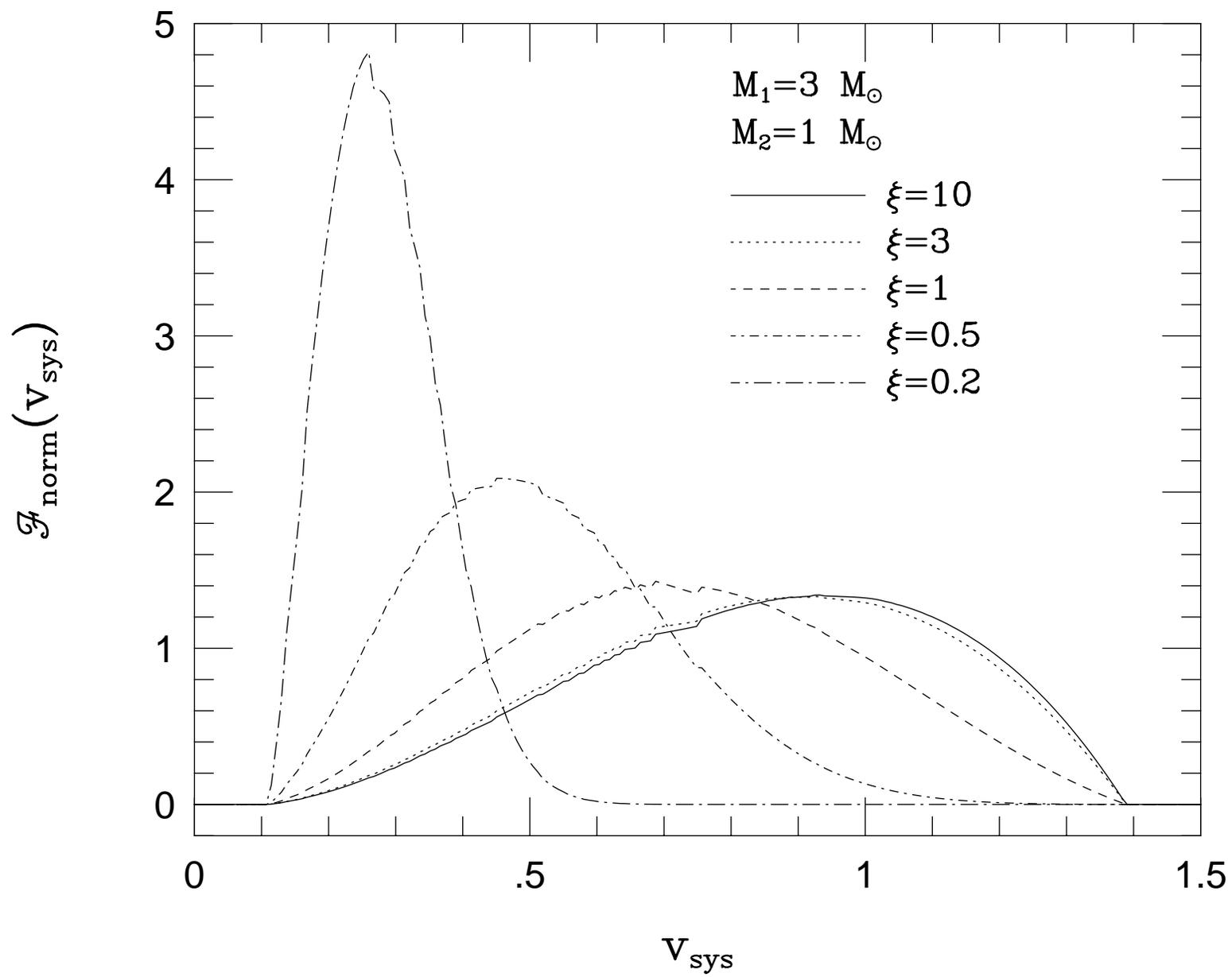
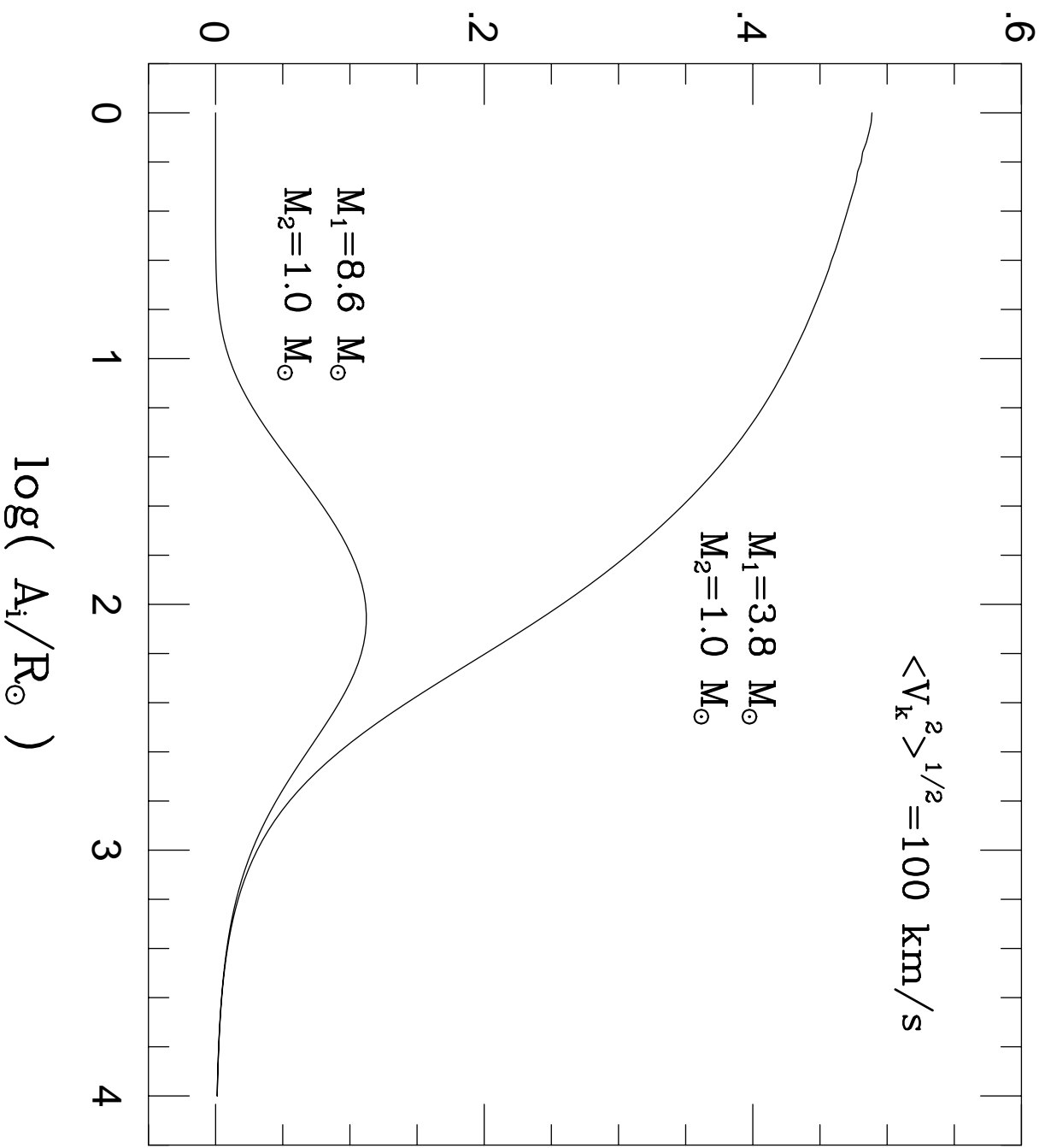
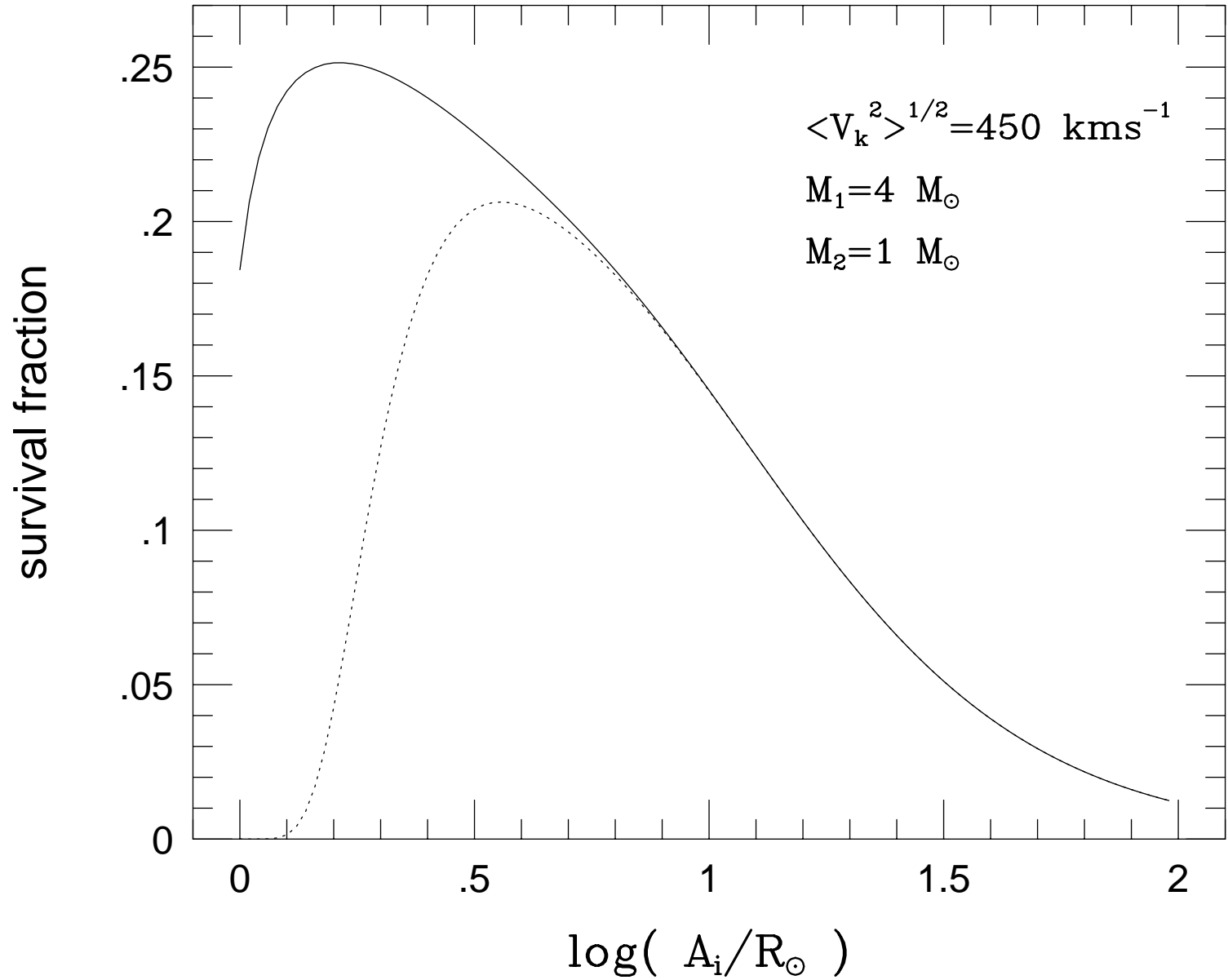
\end{document}